\pdfoutput=1
\documentclass[a4paper,12pt]{article}

\usepackage{amsfonts}
\usepackage{mathrsfs}
\usepackage{amssymb}
\usepackage{amsmath}
\usepackage{shuffle}
\usepackage{framed} 
\usepackage{subfigure}
\usepackage[medium]{titlesec}
\usepackage{bm}
\usepackage{cite}

\usepackage[normalem]{ulem}
\usepackage{extarrows}
\usepackage{slashed}
\usepackage{isodateo}
\usepackage{graphicx}
\usepackage{array}
\usepackage[dvipsnames]{xcolor}
\usepackage[bookmarksnumbered=true,bookmarksopen=true]{hyperref}
 \hypersetup{colorlinks,
             linkcolor=NavyBlue,
             citecolor=PineGreen,
             urlcolor=PineGreen}
\usepackage[hmargin=.7in,vmargin=1.1in]{geometry}
\usepackage{indentfirst}
\usepackage{booktabs}

\usepackage{bbm}

\newcommand{\FR}[2]{\displaystyle\frac{\,{#1}\,}{#2}}
\newcommand{\fr}[2]{\mbox{$\frac{\,{#1}\,}{#2}$}}
\newcommand{\n}{\nonumber}
\renewcommand{\rm}{\mathrm}

\graphicspath{{fig/}}

\def\bge{\begin{equation}}
\def\ede{\end{equation}}
\def\bga{\begin{aligned}}
\def\eda{\end{aligned}}
\def\bgb{\begin{bmatrix}}
\def\edb{\end{bmatrix}}
\def\bgp{\begin{pmatrix}}
\def\edp{\end{pmatrix}}
\def\bgm{\begin{matrix}}
\def\edm{\end{matrix}}
\def\bgs{\begin{subequations}}
\def\eds{\end{subequations}}
\newcommand{\order}[1]{\mathcal{O}({#1})}
\def\di{{\mathrm{d}}}

\def\mb{\mathbf}

\def\mc{\mathcal}

\def\pd{\partial}

\def\la{\langle}\def\ra{\rangle}

\def\to{\rightarrow}

\def\ii{\mathrm{i}}

\def\de{\delta}

\def\si{\sigma}

\def\aa{\mathsf{a}}
\def\bb{\mathsf{b}}
\def\cc{\mathsf{c}}

\DeclareFontFamily{U}{bigshuffle}{}
\DeclareFontShape{U}{bigshuffle}{m}{n}{
  <5-8> s*[1.7] shuffle7
  <8->  s*[1.7] shuffle10
}{}
\DeclareSymbolFont{BigShuffle}{U}{bigshuffle}{m}{n}
\DeclareMathSymbol{\bigshuffle}{\mathop}{BigShuffle}{"001}
\DeclareMathSymbol{\bigcshuffle}{\mathop}{BigShuffle}{"002}

\usepackage{mdframed}

\newmdenv[skipabove=0pt,%
          skipbelow=5pt,%
          leftmargin=0pt,%
          rightmargin=0pt,%
          innertopmargin=-5pt,%
          innerbottommargin=7pt,%
          innerleftmargin=2pt,%
          innerrightmargin=2pt,%
          splittopskip=0pt,%
          splitbottomskip=0pt,%
          linewidth=0pt,%
          nobreak=true]%
          {keyeqn2}

\newmdenv[backgroundcolor=gray!15,%
          skipabove=0pt,%
          skipbelow=5pt,%
          leftmargin=0pt,%
          rightmargin=0pt,%
          innertopmargin=-5pt,%
          innerbottommargin=7pt,%
          innerleftmargin=2pt,%
          innerrightmargin=2pt,%
          splittopskip=0pt,%
          splitbottomskip=0pt,%
          linewidth=0pt,%
          nobreak=true]%
          {keyeqn}

\makeatletter
\@ifpackageloaded{hyperref}%
  {\newcommand{\mylabel}[2]% #1=name, #2 = contents
    {\protected@write\@auxout{}{\string\newlabel{#1}{{#2}{\thepage}%
      {\@currentlabelname}{\@currentHref}{}}}}}%
  {\newcommand{\mylabel}[2]% #1=name, #2 = contents
    {\protected@write\@auxout{}{\string\newlabel{#1}{{#2}{\thepage}}}}}
\makeatother

\usepackage{titlesec}          
\titleformat{\section}
{\normalfont\fontsize{15}{20}\bfseries}{\thesection}{1em}{}

\newcommand{\ob}[1]{\mkern 2mu \overline{\mkern -2mu #1 \mkern -2mu}\mkern 2mu}
\newcommand{\wt}[1]{\mkern 2mu \widetilde{\mkern -2mu #1 \mkern -2mu}\mkern 2mu}

\newcommand{\fnemail}[1]{\footnote{Email: \href{mailto:#1}{\nolinkurl{#1}}}}

\newcommand{\nn}{\widetilde{\nu}}

\begin{document}

\title{\Large\textbf{On-Shell Bootstrap of Loop Inflation Correlators with Spectral Dispersion\\[2mm]}}

\author{Haoyuan Liu,$^{a\,}$\fnemail{liuhy23@mails.tsinghua.edu.cn}%
~~~~Zhehan Qin,$^{a\,}$\fnemail{qzh21@mails.tsinghua.edu.cn}%
~~~~Jiayi Wu$,^{a,b\,}$\fnemail{wu-jy22@mails.tsinghua.edu.cn}%
\\[6pt]
%~~~~
Zhong-Zhi Xianyu,$^{a,c\,}$\fnemail{zxianyu@tsinghua.edu.cn}%
~~~~Hongyu Zhang$^{\,a\,}$\fnemail{zhy21@mails.tsinghua.edu.cn}\\[5mm]
$^a\,$\normalsize{\emph{Department of Physics, Tsinghua University, Beijing 100084, China} }\\ 
$^b\,$\normalsize{\emph{Cosmology, Gravity and Astroparticle Physics Group,}}\\
      \normalsize{\emph{Center for Theoretical Physics of the Universe,}}\\
      \normalsize{\emph{Institute for Basic Science, Daejeon 34126, Korea}}\\
$^c\,$\normalsize{\emph{Peng Huanwu Center for Fundamental Theory, Hefei, Anhui 230026, China }}
}

\date{}
\maketitle

\vspace{20mm}

\begin{abstract}
\vspace{10mm}

We develop a new bootstrap strategy for cosmological correlators at loop level, which we call spectral dispersion. It is based on two conceptual observations that a correlator can be recovered from its on-shell data, also known as nonlocal signals, by analyticity up to local counterterms, and that the on-shell data for a loop process take the form of a discrete sum over quasinormal modes. Technically, our method combines the dS spectral decomposition with dispersion relations. Using this technique, we bootstrap new results in a simple and intuitive form for 3-point and 4-point correlators with 1-loop massive exchanges of scalar and vector bosons, either directly or derivatively coupled. Applications of this bootstrap technique to higher spins and higher-loop banana graphs with dS covariant dispersions but noncovariant couplings are also straightforward.

\end{abstract}

\newpage
\tableofcontents

\newpage

\section{Introduction}\label{sec_intro}

Our universe features small inhomogeneities and anisotropies at large scales, yet it is these small fluctuations that encode rich physics of the primordial era. It is widely accepted that the fast expansion during primordial inflation rapidly redshifted the microscopic quantum fluctuations of spacetime and matter to large-scale perturbations, leaving detectable imprints on the cosmological microwave background (CMB), the large-scale structure (LSS), and the 21cm line \cite{Komatsu:2001rj,Maldacena:2002vr,Babich:2004gb,Bartolo:2004if,Chen:2010xka,SPHEREx:2014bgr,Munoz:2015eqa,Meerburg:2019qqi,Planck:2018jri,Planck:2019kim,BICEP:2021xfz,Ferraro:2022cmj,Liu:2022iyy,Achucarro:2022qrl,Chaussidon:2024qni,Euclid:2024ris,Spec-S5:2025uom,SimonsObservatory:2025wwn}.

Studies of primordial fluctuations are particularly active in the recent decade under a program of ``cosmological collider (CC) physics'' \cite{Chen:2009we,Chen:2009zp,Baumann:2011nk,Chen:2012ge,Noumi:2012vr,Arkani-Hamed:2015bza,Chen:2015lza,Lee:2016vti,Meerburg:2016zdz,Chen:2016uwp,Chen:2016hrz,Baumann:2017jvh,chen:2018xck,Lu:2019tjj,Kumar:2019ebj,Liu:2019fag,Wang:2019gbi,Wang:2020ioa,Bodas:2020yho,Lu:2021wxu,Cui:2021iie,Tong:2022cdz,Reece:2022soh,Chen:2022vzh,Qin:2022lva,Craig:2024qgy,Cabass:2024wob,Sohn:2024xzd,Qin:2025xct,Bodas:2025vpb,Anbajagane:2025uro,Kumar:2025anx,Colas:2025ind,Suman:2025tpv,Jiang:2025mlm,Ferreira:2026tyj,Green:2026yev,Philcox:2026bfa,Kumar:2026ogn,Philcox:2026njr,Aoki:2026olh,Kumar:2026dih,You:2026xoq,Arundine:2026myr,Aoki:2026vbc}. This program uses the high energy of inflation, up to $\order{10^{13}\,\text{GeV}}$ in terms of Hubble scale $H$, to probe heavy particles of mass around or beyond $H$. These heavy particles could leave oscillatory features in the primordial correlation functions, known as CC signals. Theoretically, the primordial correlation functions correspond to future-boundary correlators of quantum fields in the inflationary spacetime, which is approximately a dS spacetime.  Thus, to make quantitative predictions about CC signals from a given particle-physics model, it is important to compute precisely and efficiently the correlation functions with massive exchanges.

The computation of inflation correlators is conceptually straightforward since they typically arise from a weakly coupled quantum field theory in the bulk dS, where we can perform a standard diagrammatic computation \cite{Chen:2017ryl} based on  in-in formalism \cite{Weinberg:2005vy} or the more convenient Schwinger-Keldysh path integrals \cite{Schwinger:1960qe,Feynman:1963fq,Keldysh:1964ud}. However, the computation is technically difficult due to multilayer and nested time integrals over products of special functions. Recent years have seen much progress along this direction with many new techniques developed, including differential equations \cite{Baumann:2022jpr,Arkani-Hamed:2018kmz,Baumann:2019oyu,Pimentel:2022fsc,Jazayeri:2022kjy,Qin:2022fbv,Qin:2023ejc,Aoki:2023wdc,Aoki:2024uyi,Liu:2024str,Xianyu:2025lbk,Baumann:2026atn}, the partial Mellin-Barnes representation \cite{Qin:2022lva,Qin:2022fbv,Qin:2024gtr},
the family-tree decomposition \cite{Xianyu:2023ytd,Fan:2024iek,Fan:2025scu}, dispersion relations \cite{Meltzer:2021zin,Liu:2024xyi,Werth:2024mjg,Belrhali:2026ktb,Belrhali:2026rkn}, spectral decomposition \cite{Xianyu:2022jwk,Zhang:2025nzd}, new results for a subclass of conformal scalar correlators \cite{Arkani-Hamed:2017fdk,Benincasa:2024leu,Arkani-Hamed:2023bsv,Arkani-Hamed:2023kig,Fan:2024iek,Pimentel:2026kqc}, as well as numerical approaches \cite{An:2017hlx,Wang:2021qez,Werth:2023pfl,Pinol:2023oux,Werth:2024aui}. 

Currently, the problem of computing tree-level massive correlators has been solved in terms of hypergeometric using the differential equation and the family-tree decomposition techniques \cite{Liu:2024str,Xianyu:2025lbk}. On the contrary, loops remain difficult and known results are few. Complete results exist only for 1-loop massive bubble graphs \cite{Xianyu:2022jwk,Zhang:2025nzd}. Partial results of different parts or limits of loop correlators also exist \cite{Qin:2022lva,Qin:2024gtr,Qin:2023bjk,Qin:2023nhv}. See also \cite{Marolf:2010zp,He:2024olr,Baumann:2024mvm,Cespedes:2025dnq,Bhowmick:2025mxh,Jain:2025maa,Cespedes:2025ple,Farren:2026hao,Hoory:2026gsy,Chen:2026dqp,Aoki:2026vbc} for related works. However, loop processes play very important roles in CC physics \cite{Chen:2016uwp,Chen:2016hrz,chen:2018xck,Kumar:2018jxz,Lu:2019tjj,Hook:2019zxa,Hook:2019vcn,Wang:2019gbi,Wang:2020ioa,Lu:2021gso,Cui:2021iie,Tong:2022cdz,Bodas:2025vpb,You:2026xoq,Aoki:2026vbc} and many phenomenologically important loop processes remain to be better understood quantitatively. Thus, massive inflation correlators at loop levels are now becoming a new frontier which we should push vigorously. 

\paragraph{An on-shell bootstrap program for loops}
In this work, we propose a new analytic approach to loop correlators which is a bootstrap strategy based fully on on-shell data and the analyticity of the correlators. This approach is made possible by a few observations about the analytical structure of loop correlators together with the combination of two techniques, the dispersion relation and the spectral decomposition. So, we call it spectral dispersion. Below, we explain the main idea.

A central input of our method is the analyticity. It is well known that, in Minkowski spacetime, a connected scattering amplitude is analytic everywhere except for a few poles and branch cuts. These nonanalytic behaviors typically arise due to intermediate states going on-shell, where the amplitude usually simplifies to products of smaller amplitudes. It is thus conceivable to use these nonanalyticities, or equivalently, the amplitudes for smaller processes, to bootstrap the amplitude for the full process, up to local contact terms. Such bootstrap philosophy has been very fruitful for scattering amplitudes, yielding well-known techniques such as BCFW at the tree level and general unitarity cut at the loop level \cite{Britto:2004ap,Britto:2005fq,Cutkosky:1960sp,Bern:1994zx,Bern:1994cg,Britto:2004nc}.

For inflation correlators, the situation is very similar but with an important difference. The similarity is that a correlator is also analytic everywhere in its (complex) kinematic variables, except for a few nonanalyticities, including those from intermediate states going on-shell. However, the difference is that the on-shell nonanalyticities are typically branch point \emph{at both tree and loop levels}. Heuristically, the reason is that an on-shell particle in dS is never really stable because it is diluted by the Hubble expansion. So an on-shell intermediate state cannot generate a very divergent singularity such as a pole. Instead, an on-shell propagating state typically generates (convergent) branch points, be it a one-particle state (tree) or multi-particle state (loop). In particular, as we will detail later, for a single-channel 2-to-2 process, the locations of all branch points are identical for a single-massive-particle exchange (in the left-hand side of Fig.\;\ref{fig_4ptTree}) and a double-massive-particle exchange (like in Fig.\;\ref{fd_4pt_bubble}). 

The common structure of branch points motivates a bootstrap method based on analyticity that is applicable simultaneously to tree and loop correlators. This has been achieved in \cite{Liu:2024xyi} by a \emph{line dispersion relation} which expresses the full correlator as an integral over the discontinuity of a specially chosen branch cut, up to possible local contact terms. It was shown that in \cite{Liu:2024xyi} that this discontinuity is fully determined by the on-shell data contained in the correlator, which are also known as nonlocal CC signals in the literature \cite{Tong:2021wai,Qin:2023bjk,Qin:2023nhv}.  

The nonlocal signal of an inflation correlator has received special attention in the previous studies, partly due to its intimate relation with CC phenomenology. For a tree process like Fig.\;\ref{fig_4ptTree}, the nonlocal signal arises due to the spontaneous creation of a pair of heavy particles which then undergoes on-shell resonance with external modes. This on-shell resonance manifests itself as logarithmic-scale oscillations --- a branch point --- in momentum space. Notably, a heavy scalar particle of mass $m>3H/2$ always gives rise to two branches of nonlocal signals in the form of $k_s^{2\Delta_\pm}$ where $k_s$ is the momentum it carries and $\Delta_\pm=\fr32\pm\ii\sqrt{(m/H)^2-\fr94}$ are the scaling dimensions of two late-time modes of the heavy particle. This particular pair of values of scaling dimension is a direct consequence of unitarity imposed on dS one-particle states. 

Curiously, this structure of nonlocal signal persists to all loop orders. 
In \cite{Qin:2023bjk,Qin:2023nhv}, it was proved that the nonlocal signals in inflation correlators correspond to superhorizon propagation of intermediate states going on-shell and thus satisfy a factorization theorem at the branch point very similar to the on-shell factorization of flat-space scattering amplitudes. 
Remarkably, the nonlocal signal of a loop process also breaks into contributions of late-time modes of definite scaling dimensions. 
It is just that the number of late-time modes for a loop is not 2, but countable infinity.

In \cite{Xianyu:2022jwk}, it was shown that, for a particular type of 1-loop graphs, namely the bubble graphs, the contributions from the infinite number of late-time modes can be systematically summed into a very neat form by using the dS spectral decomposition \cite{Marolf:2010zp,Loparco:2023rug}. The spectral decomposition is the dS counterpart of Källén-Lehmann representation in Minkowski spacetime, which represents a bubble loop, or more generally, any dS-covariant two-point function, as a spectral integral over the mass of a massive propagator, weighted by a spectral function. Although the spectral integral is an integral over mass, the result for the nonlocal signal features a discrete sum over scaling dimensions, which can be traced to the meromorphicity of a general spectral function in dS \cite{Mirbabayi:2022gnl}. 

The results reviewed above let us conjecture a shortcut to the full result of a 1-loop, which consists of two steps:
\begin{enumerate}
  \item Perhaps, there exists a ``line dispersion relation,'' similar to the one proposed in \cite{Liu:2024xyi}, but applicable to an individual late-time mode with fixed scaling dimension $\Delta$ rather than a particle of fixed mass. Were this true, it would allow us to find a direct one-to-one map between the full correlator $\mb{I}^\Delta$ exchanging a single mode $\Delta$ with its nonlocal signal $\mb{I}_\text{NS}^\Delta$. 
  
  \item For a bubble loop graph $\mc{J}$, we can write down its nonlocal signal $\mc{J}_\text{NS}$ as a discrete sum of nonlocal signals of different modes: $\mc{J}_\text{NS}=\sum_{n,\Delta}\mc{C}_{n,\Delta}\mb{I}_\text{NS}^{\Delta+2n}$ with some spectral coefficients $\mc{C}_{n,\Delta}$ fully determined by spectral decomposition. Now, since the bubble loop graph $\mathcal{J}$ shares identical analytic structure with a tree graph, we would further conjecture that its full result can simply be written as $\mathcal{J}=\sum_{n,\Delta}\mc{C}_{n,\Delta}\mb{I}^{\Delta+2n}$. 
\end{enumerate}

If we can show that the above two conjectures are true, then we will be able to directly write down the full result for a 1-loop bubble graph from the (known) tree correlator exchanging one single mode and the (known) spectral coefficients. In this work, we will show that this is indeed the case, except for local terms contributed from a contact interaction. As will be detailed later, these terms should be determined by a renormalization condition for the 1-loop process and thus are never expected to be fixed by a dispersive calculation based only on analyticity and on-shell data. 

We call the shortcut \emph{spectral dispersion} as it exploits both the spectral decomposition and the dispersion relation as useful bootstrap techniques. Conceptually, it can be viewed as a dS counterpart of the well-developed general unitarity cuts for flat-space amplitudes.

\begin{figure}[t]
  \centering 
  \includegraphics[width=0.99\textwidth]{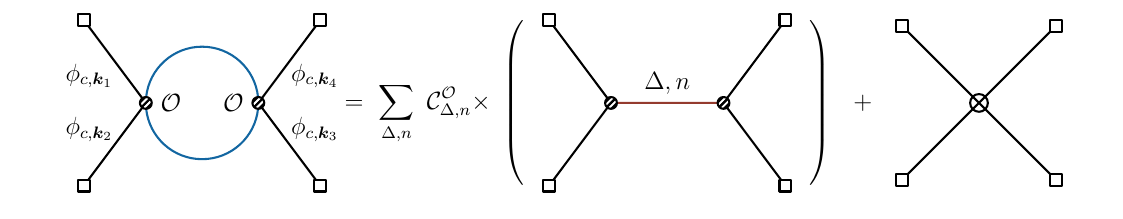} 
  \caption{The spectral dispersion of the 4-point correlator of the conformal scalar $\phi_c$ with massive 1-loop bubble exchange through the operator $\mathcal{O}$.
  }
\label{fd_4pt_bubble_dispersion}
\end{figure}

Armed with this new tool, we bootstrap a few 1-loop bubble processes mediating both massive scalars and vector bosons, either directly or derivatively coupled, which are schematically shown in Fig.\;\ref{fd_4pt_bubble_dispersion} (and also in Fig.\;\ref{fd_3pt_bubble} for 3-point functions). Similar results have been computed with a purely spectral method in \cite{Xianyu:2022jwk,Zhang:2025nzd}, but the results here feature much simpler form and show explicitly the spectral structure of these loop processes. Given that the spectral decomposition can be done for a pair of particles of arbitrary masses and spins, application of our method to 1-loop processes involving more general spins and couplings are also straightforward. The same is also true for banana loops, where we can apply spectral decomposition iteratively. Also important is that our method is applicable to  bubble and banana loops of boost-breaking couplings. We will comment on this point in more detail in Sec.\;\ref{sec_spectral}.

\paragraph{Outline of this work}
The rest of this work is structured as follows. 
We first review the two major technical ingredients of our methods, the dispersion relation in Sec.\;\ref{sec_dispersion_tree} and the spectral decomposition in Sec.\;\ref{sec_spectral}, with a few technical details collected in App.\;\ref{appd_spectral}.
The new result here is that we can develop line dispersion relations for a correlator mediating a single late-time mode with scaling dimension $\Delta$, as shown in (\ref{eq_QNMLineDispersion}). We also put an emphasis on the fact that the nonlocal signal for a loop process can be expressed as a discrete sum over nonlocal signals of modes, as shown in (\ref{eq_NS_spectral_sum}). 

We then spell out our bootstrap strategy more explicitly in Sec.\;\ref{sec_dispersion_method} and apply it to a 3-point inflaton correlator in Sec.\;\ref{sec_dispersion_method_3pt} and a few 4-point correlators of conformal scalars with intermediate massive bubble exchanges in Sec.\;\ref{sec_dispersion_result}. Applications of our results to more realistic inflation correlators are then discussed in Sec.\;\ref{sec_pheno} for both trispectrum and bispectrum. We conclude in Sec.\;\ref{sec_conclusion}.

\paragraph{Notations and conventions} We work in the inflation patch of the dS spacetime where the metric reads $\di s^2=a^2(\tau)(-\di\tau^2+\di\bm x^2)$, with $\tau\in(-\infty,0)$ the conformal time and $\bm x\in\mathbb{R}^3$ the spatial comoving coordinates. The scale factor is $a(\tau)=-1/(H\tau)$ and the Hubble parameter $H$ is set to unity, $H=1$, throughout this work. We use bold italic letters such as $\bm k$ to denote 3-momenta and the corresponding italic letter $k\equiv |\bm k|$ to denote its magnitude. We also use a shorthand notation $k_{12}\equiv k_1+k_2$ to represent sums of several indexed quantities. Besides, some frequently used special functions and symbols are listed in App.\;\ref{appd_function}.

\section{Dispersive Bootstrap in a Nutshell}\label{sec_dispersion_tree}

\begin{figure}[t]
\centering 
\includegraphics[width=0.9\textwidth]{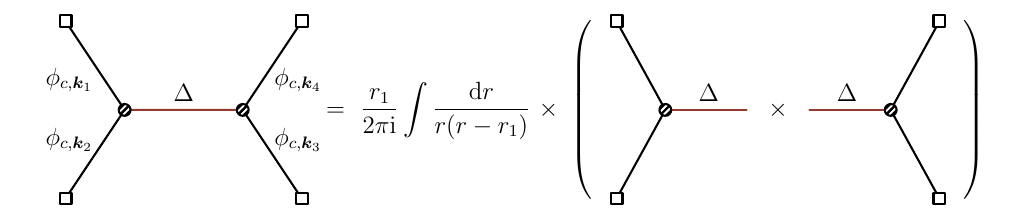} 
\caption{The four-point correlator of the conformal scalar $\phi_c$ mediated by a massive scalar field $\sigma$ at the tree level. }
\label{fig_4ptTree}
\end{figure}

The first ingredient for our spectral dispersion method is the dispersion relation. In this section, we briefly review the dispersive bootstrap at the tree level. This method was proposed and used to bootstrap massive inflation correlators in \cite{Liu:2024xyi}, which we closely follow. The new result here is that the dispersion relations hold not only for graphs mediating a particle of definite mass, but also for graphs mediating a ``quasinormal mode'' with definite scaling dimension, summarized in \eqref{eq_QNMLineDispersion} and in Fig.\;\ref{fig_4ptTree}. This quasinormal-mode-level dispersion relation will be a building block for our spectral dispersion.  

To be specific, let us consider the simplest four-point correlator of conformal scalars $\phi_c$, generated from the $s$-channel exchange of a massive field $\sigma$ with mass $m_\si$, shown in Fig.\;\ref{fig_4ptTree}. We assume that the process comes from the following interaction:
\begin{align}
\label{eq_TreeInt}
  \Delta\mathcal{L}=-\FR{1}{2}a^4\phi_c^2\sigma.
\end{align}
Throughout this paper we omit coupling constants just for simplicity. 
More realistic models with various interactions where external modes are inflaton fluctuations can be obtained by acting weight-shifting operators (taking derivatives with respect to kinematic variables) \cite{Arkani-Hamed:2015bza,Arkani-Hamed:2018kmz,Baumann:2019oyu}. 
 
Following the standard Schwinger-Keldysh formalism \cite{Schwinger:1960qe,Feynman:1963fq,Keldysh:1964ud,Weinberg:2005vy} (see  \cite{Chen:2017ryl} for a review), the four-point correlator reads:
\begin{align}
 \label{eq_4ptTreeCorrelator}
 \la\phi_{c,\bm{k}_1}\phi_{c,\bm{k}_2}\phi_{c,\bm{k}_3}\phi_{c,\bm{k}_4}\ra' =&\,\sum_{\aa,\bb = \pm}(-\aa\bb)\int_{-\infty}^0\FR{\di\tau_1}{\tau_1^4}\FR{\di\tau_2}{\tau_2^4}\,K_\aa(k_1,\tau_1)K_\aa(k_2,\tau_1)\n\\
 &\times K_\bb(k_3,\tau_2)K_\bb(k_4,\tau_2)\times D^{\nn}_{\aa\bb}(k_s;\tau_1,\tau_2).
\end{align} 
It is understood that $\la\cdots\ra'$ means the momentum conservation factor $(2\pi)^3\de^{(3)}(\bm k_1+\bm k_2+\bm k_3+\bm k_4)$ dropped off and is not to be confused with a conformal-time derivative. Thanks to the Weyl symmetry of a free conformal scalar, the bulk-to-boundary propagators for the conformal scalar $\phi_c$ has a simple form:  
\begin{equation}\label{eq_conformal_propagator}
K_\pm(k,\tau) = \FR{\tau\tau_f}{2k}e^{\pm\ii k\tau},
\end{equation} 
with $\tau_f$ the (irrelevant) late-time cutoff, while the bulk propagators for $\si$ are rather complicated because of the spacetime curvature: 
\begin{align}\label{eq_propagator_si}
  &D^{\nn}_{-+}(k;\tau_1,\tau_2) = u(k,\tau_1)u^*(k,\tau_2),\qquad D^{\nn}_{+-}(k;\tau_1,\tau_2) = u^*(k,\tau_1)u(k,\tau_2),\\
  &D^{\nn}_{\pm\pm}(k;\tau_1,\tau_2) = D^{\nn}_{\mp\pm}(k;\tau_1,\tau_2) \theta(\tau_1-\tau_2) + D^{\nn}_{\pm\mp}(k;\tau_1,\tau_2) \theta(\tau_2-\tau_1),
\end{align} 
where $u(k,\tau)$ is the mode function for $\si$ field, namely, a properly normalized solution to the massive Klein-Gordon equation with the Bunch-Davies initial condition:
\begin{equation}
u(k,\tau) = \FR{\sqrt\pi}{2}e^{-\pi\wt\nu/2}(-\tau)^{3/2}\mathrm H_{\ii\wt\nu}^{(1)}(-k\tau),\qquad \wt\nu\equiv \sqrt{m_\sigma^2-\FR94}.
\end{equation}

The four-point correlator (\ref{eq_4ptTreeCorrelator}) can be made dimensionless by taking out a purely kinematic dimensionful factor:
\begin{equation}\label{eq_tree_seed_from_correlator}
  \la\phi_{c,\bm{k}_1}\phi_{c,\bm{k}_2}\phi_{c,\bm{k}_3}\phi_{c,\bm{k}_4}\ra' = \FR{\tau_f^4}{16k_1k_2k_3k_4k_s}\mathcal{I}(r_1,x),
\end{equation}
The resulting \emph{tree seed integral} $\mathcal I$ is a dimensionless function of two independent momentum ratios, which can be chosen in many ways. Here, we choose $r_1$ and $x$ which are defined as:
\begin{align}\label{eq_def_r_and_x}
  r_1\equiv\FR{k_s}{k_{12}},\qquad x\equiv\FR{k_{34}}{k_{12}}.
\end{align}
The definition of the seed integral $\mathcal I$ follows from (\ref{eq_4ptTreeCorrelator}) and (\ref{eq_tree_seed_from_correlator}): 
\begin{equation}\label{eq_TreeSeedDef}
  \mathcal I \Big( \FR{k_s}{k_{12}},\FR{k_{34}}{k_{12}} \Big) \equiv k_s \sum_{\aa,\bb=\pm}(-\aa\bb)\int_{-\infty}^0 \FR{\di\tau_1}{\tau_1^2}\FR{\di\tau_2}{\tau_2^2}\,D_{\aa\bb}^{\nn}(k_s;\tau_1,\tau_2)e^{\ii\aa k_{12}\tau_1+\ii\bb k_{34}\tau_2}.
\end{equation} 
Below, we will review the analytical structure of the tree seed integral \eqref{eq_TreeSeedDef} and use it as the bootstrap resource for a dispersive calculation of \eqref{eq_TreeSeedDef} itself.

\begin{figure}[t]
\centering 
\includegraphics[width=0.5\textwidth]{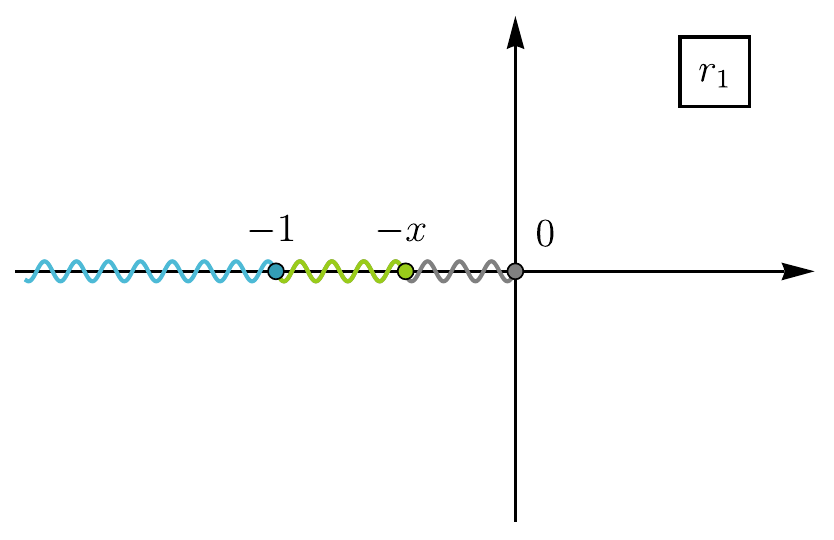} 
\caption{The analytical structure of $\mathcal{I}(r_1,x)$ on $r_1$ plane. The wavy lines represent branch cuts, and the dots represent branch points. Here we assume $x=k_{34}/k_{12}\in(0,1)$. The branch cut extends to the branch point $r_1\to-\infty$, which is not shown in this figure. See also Fig.\;10 of \cite{Liu:2024xyi}. }
\label{fig_branch}
\end{figure}

\paragraph{Nonanalyticity}
We now review the singularity structure of the tree seed integral $\mathcal I$, which is of central importance for forming dispersion integrals. 
Since $\mathcal I$ is a function of two variables, we fix one of them in the physical region and show the analytical structure of $\mc I$ on the plane of one complex variable.

For simplicity, we choose to fix $x$ and study the analytical structures on the complex $r_1$ plane.\footnote{As explained in \cite{Liu:2024xyi}, constructing a dispersion integral on the $r_1$ plane is more economic than on the $x$ plane.} \emph{A priori}, the entire physical region for $x$ is the entire real positive axis, since $k_{12}$ and $k_{34}$ can take any real values independently. However, the exchange symmetry $k_{12}\leftrightarrow k_{34}$ of the correlator induces the following identity for the seed integral:
\begin{equation}
\label{eq_trans}
\mathcal I(r_1,x) = \mathcal I\Big(\FR{r_1}x,\FR{1}{x}\Big),
\end{equation}
which allows us to restrict $x\in(0,1]$, because otherwise we can always use \eqref{eq_trans} to change the variable so that $x$ stays in $(0,1]$. With $x$ fixed, we can regard $\mathcal I$ as a function of $r_1$ only and write $\mathcal I(r_1) \equiv \mathcal I(r_1,x)$ for simplicity.

The analytical structure of $\mathcal I (r_1)$ has been analyzed in detail in \cite{Liu:2024xyi}. 
In short, the function $\mathcal I (r_1)$ is fully analytic on the entire $r_1$ plane, except for 4 branch points at $r_1=0$, $-x$, $-1,$ and $\infty$. 
As a result, the branch cut can be chosen to be on the entire negative real axis, and the discontinuity across this branch cuts exhibits discontinuity at $r_1=-1$ and $r_1=-x$; 
See Fig.\;\ref{fig_branch}.

The physical meanings of the four branch points are clear. First, the branch point $r_1=0$ corresponds to the squeezed limit $k_s\to 0$ (also known as the collapsed limit). Second, the branch points $r_1\to-1$ and $r_1\to-x$ correspond to the partial-energy ``poles'' at $k_s+k_{12} \to 0$ and $k_s+k_{34}\to0$, respectively. Finally, the branch point $r_1\to-\infty$ corresponds to the total energy ``pole'' at $k_{1234}\to0$.\footnote{To see that $r_1\to -\infty$ with fixed $x$ is the total-energy limit, notice that $r_1=k_s/k_{12}\to-\infty$ can be reached by sending $k_{12}\to0$ with fixed $k_s$. So, with $x\in(0,1]$ fixed, the  total energy $k_{12}+k_{34}=(1+x)k_{12}\to0$.} 

It is well understood that the nonanalyticity around $r_1\to 0$ with fixed $x$ (or, equivalently, around the squeezed limit $k_s\to 0$ with fixed $k_{12}$ and $k_{34}$) is characterized by the \emph{nonlocal signal} $\mathcal I_\text{NS}$ \cite{Qin:2023bjk,Qin:2023nhv} due to on-shell particle productions \cite{Tong:2021wai}:
\begin{equation}
\label{eq_TreeNonlocal}
\mathcal I (r_1) \supset \mathcal I_\text{NS}(r_1) = \mathbf{I}^{\Delta_{+}}_{\text{NS}}(r_1)+ \mathbf{I}^{\Delta_{-}}_{\text{NS}}(r_1),
\qquad \Delta_\pm = \FR32\pm \ii\wt\nu,
\end{equation}
where we have defined 
\begin{equation}\label{eq_tree_NS_QNmode}
 \mathbf{I}^{\Delta}_{\text{NS}}(r_1)\equiv\FR{\sec(\pi\Delta)+1}{4\cos(\pi\Delta)}\,x^{1-\Delta}\,r_1^{2\Delta-2}\,{}_2\mathcal F_1\biggl[\bgm
 \fr{\Delta-1}2,\fr\Delta2 \\ \Delta-\fr12\edm\bigg|r_1^2\biggr]{}_2\mathcal F_1\biggl[\bgm
 \fr{\Delta-1}2,\fr\Delta2 \\ \Delta-\fr12\edm\bigg|\FR{r_1^2}{x^2}\biggr],
\end{equation}
with ${}_2\mathcal{F}_1[\cdots]$ the dressed hypergeometric function defined in App.\;\ref{appd_function}.
The decomposition of $\mathcal I_\text{NS}$ into $\mb I_\text{NS}^{\Delta_+}$ and $\mb I_\text{NS}^{\Delta_-}$ is intuitive:
When approaching the future boundary, the bulk operator $\si$ splits into a pair of primary operators $\sigma_\pm$ at the boundary with scaling dimensions $\Delta_\pm$, together with their descendants:
\begin{equation}
\lim_{\tau\to 0} \sigma(\tau,\bm x) \sim \sigma_+(\bm x)(-\tau)^{\Delta_+} + \sigma_-(\bm x)(-\tau)^{\Delta_-} + \text{descendants}.
\end{equation}
By counting the power of $k_s$ (encoded in the power of $r_1$), we identify $\mb I_\text{NS}^{\Delta_\pm}$ as the contribution of these two operator families after the Fourier transform $\bm x\to \bm k_s$.

Since the nonanalytical behavior of the tree seed integral $\mathcal{I}(r_1)$ at $r_1=0$ is fully captured by nonlocal signal, we immediately know that the nonlocal signal itself fully provides the discontinuity of $\mathcal{I}(r_1)$ around $r_1=0$:
\begin{equation}
\lim_{r_1\to 0}\mathop{\text{Disc}}\mathcal{I}(r_1)=\lim_{r_1\to 0}\mathop{\text{Disc}}\Big[ \mathcal{I}_\text{NS}(r_1)\Big] \theta(-r_1),
\end{equation}
where the step function $\theta(-r_1)$ indicates that the branch cut is set along the negative real axis, and we adopt the following common definition of discontinuity:
\begin{align}
  \mathop{\text{Disc}}f(z)\equiv \lim_{\epsilon\to0}\big[f(z+\ii\epsilon)-f(z-\ii\epsilon)\big],\qquad z\in \mathbb R.
\end{align}
A more nontrivial relation was derived in \cite{Liu:2024xyi} which relates the discontinuities of $\mathcal I (r_1)$ and the nonlocal signal on the whole real axis, not only around the squeezed limit:
\begin{align}
\label{eq_TreeDiscRelation}
    \mathop{\text{Disc}}\mathcal{I}(r_1)=\mathop{\text{Disc}}\Big[\mathcal{I}_\text{NS}(r_1)-\mathcal{I}_\text{NS}(-r_1)\Big]\theta(-r_1),\qquad r_1\in \mathbb R.
\end{align}
We emphasize that the relation \eqref{eq_TreeDiscRelation} depends on the form of the interaction \eqref{eq_TreeInt}.\footnote{For correlators with other interactions, similar relations can be derived as well; See \cite{Liu:2024xyi}.} 

The relation \eqref{eq_TreeDiscRelation} together with the explicit expression for the nonlocal signal \eqref{eq_TreeNonlocal} provides us the necessary bootstrap ingredient to construct the dispersion integral for the tree seed integral.

\begin{figure}[t]
\centering 
\includegraphics[width=0.5\textwidth]{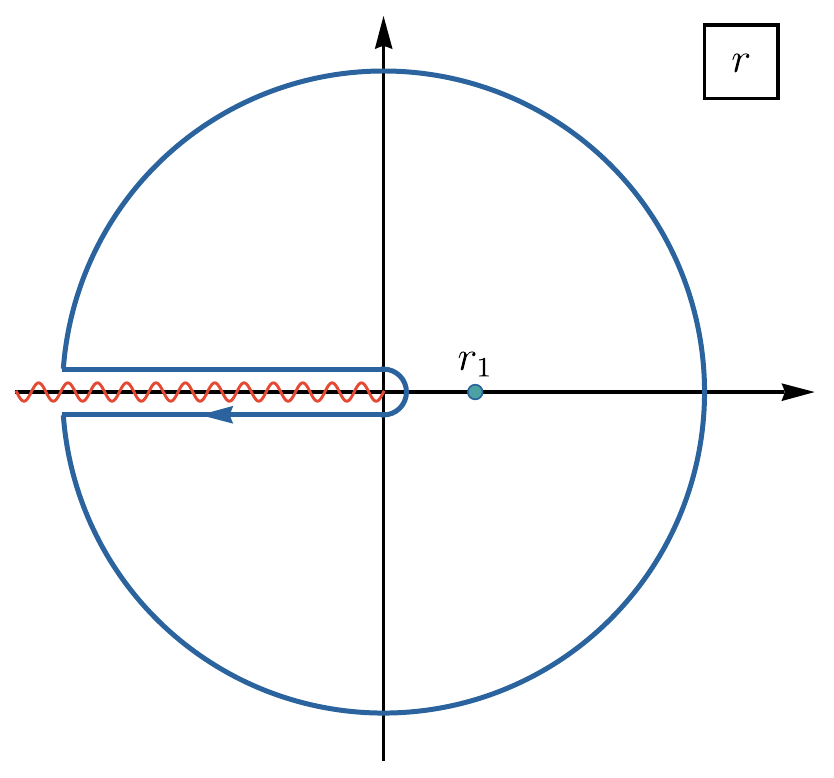} 
\caption{The dispersion relation of $\mathcal{I}$. The red wavy line represents the branch cut, and the directed blue line represents the contour of Cauchy integral, which is used to obtain the value of $\mathcal{I}$ at $r_1$. Here the integration contour has been deformed so that only integral along the branch cut is non-vanishing. }
\label{fig_dispersion}
\end{figure}

\paragraph{Line dispersion}
The dispersion relation is useful to recover (the analytic part of) a complex function from its nonanalyticities. In \cite{Liu:2024xyi}, two dispersion relations were constructed for the tree-level correlator \eqref{eq_TreeSeedDef}. One of them is called \emph{line dispersion relation}, as it utilizes the nonanalyticities on complex momentum $k_s$ flowing through the exchanged line, or equivalently, on complex $r_1$ with fixed $x$. This is the  dispersion relation we adopt in this work.

For nonexperts, the main point of a dispersion relation is that any function (potentially with poles or branch cuts, but analytic elsewhere) that decays fast enough at infinity can be obtained by performing a Cauchy integral along an appropriate contour, and our tree seed integral \eqref{eq_TreeSeedDef} is no exception: 
\begin{align}
  \mathcal{I}(r_1)=\FR{1}{2\pi\ii}\oint_C\di r\FR{\mathcal{I}(r)}{r-r_1}.
\end{align}
Here $C$ denotes an anticlockwise contour, within which there is only a simple pole at $r_1$.
Then with the deformation of integration contour shown in Fig.\;\ref{fig_dispersion}, non-vanishing contributions only come from the discontinuities along the branch cut. Using (\ref{eq_TreeDiscRelation}) for the discontinuity, we see that the seed integral satisfies the following dispersion relation:
\begin{align}\label{eq_dispersion_tree}
  \mathcal{I}(r_1)=&~\FR{r_1}{2\pi\ii}\int_{-\infty}^0\FR{\di r}{r(r-r_1)}\mathop{\text{Disc}}\Big[ \mathcal{I}_\text{NS}(r)-\mathcal{I}_\text{NS}(-r)\Big]\n\\
  =&\sum_{\Delta=\Delta_\pm}\FR{r_1}{2\pi\ii}\int_{-\infty}^0\FR{\di r}{r(r-r_1)}\mathop{\text{Disc}}\Big[ \mathbf{I}^{\Delta}_\text{NS}(r)-\mathbf{I}^{\Delta}_\text{NS}(-r)\Big].
\end{align}
Here, an extra factor of $r_1/r$ is introduced in the integrand to make it decrease fast enough at infinity so that the big circle in Fig.\;\ref{fig_dispersion} does not contribute to the final result, which is conventionally called a ``subtraction.'' The line dispersion relation (\ref{eq_dispersion_tree}) will be frequently used in this work when dealing with loop correlators.

The dispersion integral \eqref{eq_dispersion_tree} was carried out analytically in \cite{Liu:2024xyi}. Here we quote the result: 
\begin{align}
\label{eq_treeResult}
  \mathcal{I}(r_1) 
   = \sum_{\Delta=\Delta_\pm}\Big[\mathbf{I}^{\Delta}_{\text{NS}}(r_1)+\mathbf{I}^{\Delta}_{\text{LS}}(r_1)+\mathbf{I}^{\Delta}_{\text{BG}}(r_1)\Big],
\end{align}
where $\mathbf{I}^{\Delta_\pm}_{\text{NS}}$, given in (\ref{eq_tree_NS_QNmode}), is the nonlocal signal characterizing the nonanalyticity around $r_1=0$ with fixed $x$. On the other hand, $\mathbf{I}_{\text{LS}}^{\Delta_\pm}$ is the \emph{local signal} $\mathcal{I}_{\text{LS}}$ which characterizes the nonanalyticity at the limit $x\to0$ with fixed $r_1$ and has the following expression: 
\begin{align}\label{eq_Treelocal}
  \mathcal{I}_{\text{LS}}(r_1,x)= \mathbf{I}_{\text{LS}}^{\Delta_+}(r_1,x)+\mathbf{I}_{\text{LS}}^{\Delta_-}(r_1,x),
\end{align}
where
\begin{align}\label{eq_tree_LS_QNmode}
  \mathbf{I}^{\Delta}_{\text{LS}}(r_1,x)=-\FR{1}{2}\sec(\pi\Delta)^2\cos(\pi\Delta/2)^2\,r_1\,x^{\Delta-2}\,{}_2\mathcal{F}_1\bigg[\bgm\fr{\Delta-1}{2},\fr{\Delta}{2}\\\Delta-\fr{1}{2}\edm\bigg|r_1^2\bigg]{}_2\mathcal{F}_1\bigg[\bgm\fr{\ob\Delta-1}{2},\fr{\ob\Delta}{2}\\ \ob\Delta-\fr{1}{2}\edm\bigg|\FR{r_1^2}{x^2}\bigg],
\end{align}
where we define $\ob\Delta\equiv 3-\Delta$. 
Finally, the term $\mathbf{I}_{\text{BG}}^{\Delta_\pm}$ gives rise to the \emph{background} $\mathcal{I}_{\text{BG}}$, which is regular at the above two limits and can be expressed as: 
\begin{align}
  \mathcal{I}_{\text{BG}}(r_1,x)= \mathbf{I}^{\Delta_+}_{\text{BG}}(r_1,x)+\mathbf{I}^{\Delta_-}_{\text{BG}}(r_1,x),
\end{align}
where
\begin{align}\label{eq_tree_BG_QNmode}
  \mathbf{I}^{\Delta}_{\text{BG}}(r_1,x)=\sum_{n=0}^{\infty}\FR{\sec(\pi\Delta)+1}{(\Delta+n-1)(\ob\Delta+n-1)}\,r_1\,(-x)^n\,{}_3\text{F}_2\bigg[\bgm1,\fr{n}{2}+\fr{1}{2},\fr{n}{2}+1\\\fr{\Delta+n+1}{2} ,\fr{\ob\Delta+n+1}{2}\edm\bigg|r_1^2\bigg],
\end{align} 
Similar to the nonlocal signal, the separation of the local signal $\mathcal{I}_{\text{LS}}$ into $\mathbf{I}^{\Delta_\pm}_{\text{LS}}$ is a consequence of the two late-time modes with definite scaling dimensions $\Delta_\pm$. Although it is not obvious that the separation of the background into $\mathbf{I}^{\Delta_\pm}_{\text{BG}}$ can be interpreted similarly, the separation of all three pieces into two parts in (\ref{eq_treeResult}) motivates us to rewrite the dispersion integral in a way that holds separately for each scaling dimension:
\begin{keyeqn}
\begin{align}
\label{eq_QNMLineDispersion}
  \mathbf{I}^{\Delta}(r_1,x)=\FR{r_1}{2\pi\ii}\int_{-\infty}^0\FR{\di r}{r(r-r_1)}\mathop{\text{Disc}}\limits_{r}\Big[ \mathbf{I}_\text{NS}^{\Delta}(r,x)-\mathbf{I}^{\Delta}_\text{NS}(-r,x)\Big],
\end{align}
\end{keyeqn}
where $\mathbf{I}^{\Delta}\equiv\mathbf{I}^{\Delta}_{\text{NS}}+\mathbf{I}^{\Delta}_{\text{LS}}+\mathbf{I}^{\Delta}_{\text{BG}}$. This equation will play a major role in our following discussion of spectral dispersion. It shows that the line dispersion relation can be applied to a tree graph mediating a massive mode with definite scaling dimension $\Delta$. In the following, we will apply a more general version of (\ref{eq_QNMLineDispersion}) where we allow the scaling dimension $\Delta$ to take general complex values, beyond the values $\Delta=\fr32\pm\ii\wt\nu$ required for dS one-particle states. With slight abuse of terminology, we will call such a mode with definite complex $\Delta$ a \emph{quasinormal mode}.

\section{Spectral Decomposition in a Nutshell}\label{sec_spectral}

\begin{figure}[t]
  \centering 
  \includegraphics[width=0.3\textwidth]{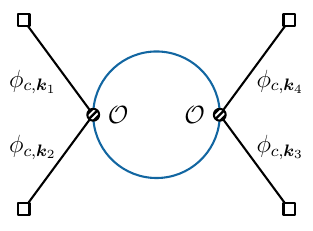} 
  \caption{The 4-point correlator of the conformal scalar $\phi_c$. Here the exchanged fields can be massive scalar $\sigma$ (which can couple to $\phi_c$ directly or derivatively) or massive vector $A_{\mu}$.}
\label{fd_4pt_bubble}
\end{figure}

In this section, we review the second ingredient of our bootstrap program, the spectral decomposition. This technique exploits the dS isometries to rewrite any dS-covariant two-point function as an integral of a tree-level massive propagator over its mass, weighted by a spectral function. This technique is the dS counterpart of Källén-Lehmann representation in flat spacetime, and has been frequently used in the study of inflation correlators in recent years \cite{Marolf:2010nz,Xianyu:2022jwk,Loparco:2023rug,Zhang:2025nzd,Grafe:2026qsm}. 

As shown in previous works, dS spectral decomposition is particularly suitable for computing correlators at loop level. In particular, we will consider 1-loop correlators with bubble topology, where the exchanged fields are massive scalars or vectors. (See Fig.\;\ref{fd_4pt_bubble}.) 
In the simplest case, the bubble correlator is mediated by a pair of massive scalar fields $\si$, and $\si$ is directly coupled to the external conformal modes:
\begin{equation}\label{eq_int_si}
    \Delta\mathcal{L}=-\FR{1}4a^4\phi_c^2\sigma^2.
\end{equation}
The bubble correlator from this type of direct coupling has been studied in \cite{Xianyu:2022jwk}. In addition, we also consider the following derivative coupling: 
\begin{equation}\label{eq_int_patsi}
    \Delta\mathcal{L}=-\FR{1}{4}a^4\phi_c^2(\nabla\sigma)^2, 
\end{equation}
where $(\nabla\sigma)^2\equiv g^{\mu\nu}\partial_{\mu}\sigma\partial_{\nu}\sigma$. This coupling may not be of immediate interest for phenomenological applications, but it serves as an important technical step for our studying spinning exchanges.

For spinning exchanges, we consider a pair of massive vector bosons $A_{\mu}$ coupled directly to external conformal scalars:
\begin{equation}\label{eq_int_A}
    \Delta\mathcal{L}=-\FR{1}{4}a^4\phi_c^2A^2, 
\end{equation}
where $A^2\equiv g^{\mu\nu}A_\mu A_\nu$. Using the diagrammatic rules, the $s$-channel bubble correlators arising from all three couplings above can be written in the following unified form:
\begin{align}\label{eq_correlator_4ptbubble}
    \langle\phi_{\bm{k}_1}\phi_{\bm{k}_2}\phi_{\bm{k}_3}\phi_{\bm{k}_4}\rangle_{\mathcal{O}}'=&\ \FR{1}{2}\sum_{\mathsf{a},\mathsf{b}=\pm}(-\mathsf{ab})\int_{-\infty}^{0}\FR{\di\tau_1}{(-\tau_1)^4}\FR{\di\tau_2}{(-\tau_2)^4}K_{\mathsf{a}}(k_1,\tau_1)K_{\mathsf{a}}(k_2,\tau_1)\n \\
    &\times K_{\mathsf{b}}(k_3,\tau_2)K_{\mathsf{b}}(k_4,\tau_2)\times\mathcal{Q}^{\mathcal{O}}_{\mathsf{ab}}\big(k_s;\tau_1,\tau_2\big),
\end{align} 
where the bulk-to-boundary propagator $K_{\pm}$ of the conformal scalar $\phi_c$ is given in (\ref{eq_conformal_propagator}), 
and $\mathcal{Q}^{\mathcal{O}}_{\mathsf{ab}}$ is the 1-loop momentum integral of bulk propagators whose explicit expression will be given later.
Here, we use $\mathcal{O}$ to represent various interactions, with $\mathcal{O}=\si^2,(\nabla\si)^2,A^2$ indicating contribution from interaction (\ref{eq_int_si}), (\ref{eq_int_patsi}), and (\ref{eq_int_A}), respectively.
Similar to the tree-level case, we define a dimensionless \emph{loop seed integral} $\mathcal{J}^{\mathcal{O}}$ from the correlator (\ref{eq_correlator_4ptbubble}) which can again be chosen as a function of $r_1\equiv k_s/k_{12}$ and $x\equiv k_{34}/k_{12}$:
\begin{align}\label{eq_LoopSeedDef}
    \mathcal{J}^{\mathcal{O}}\Big( \FR{k_s}{k_{12}},\FR{k_{34}}{k_{12}} \Big)\equiv \FR{k_s}{2}\!\sum_{\aa,\bb=\pm}(-\aa\bb)\int_{-\infty}^0\!\FR{\di\tau_1}{\tau_1^2}\FR{\di\tau_2}{\tau_2^2}\mathcal{Q}^{\mathcal{O}}_{\aa\bb}\big(k_s;\tau_1,\tau_2\big)e^{\ii\aa k_{12}\tau_1+\ii \bb k_{34}\tau_2}.
\end{align}
With the definition of $\mathcal{J}^{\mathcal{O}}$, the 1-loop correlator can be expressed as
\begin{align}
    \langle\phi_{\bm{k}_1}\phi_{\bm{k}_2}\phi_{\bm{k}_3}\phi_{\bm{k}_4}\rangle'_{\mathcal{O}}=\FR{\tau_f^4}{16k_1k_2k_3k_4k_s}\mathcal{J}^{\mathcal{O}}(r_1,x).
\end{align}
The explicit expression of 1-loop momentum integral $\mathcal{Q}^{\mathcal{O}}_{\aa\bb}$ depends on the interaction. For directly-coupled-scalar case (\ref{eq_int_si}), we have
\begin{align}\label{eq_Q_si}
    \mathcal{Q}^{\si^2}_{\mathsf{ab}}(k_s;\tau_1,\tau_2)=\int\FR{\di^3\bm{q}}{(2\pi)^3}D^{\nn}_{\mathsf{ab}}(q;\tau_1,\tau_2)D^{\nn}_{\mathsf{ab}}(|\bm{k}_s-\bm{q}|;\tau_1,\tau_2),
\end{align} 
where the bulk propagator $D_{\aa\bb}^{\nn}$ is given in (\ref{eq_propagator_si}). For derivatively-coupled scalar loop generated by (\ref{eq_int_patsi}), the bulk propagator is associated with $D^\si_{\aa\bb}$ through taking derivatives in position space, and correspondingly the 1-loop momentum integral can be expressed as: 
\begin{align}\label{1L4PNaSiNaSiMoIn}
    \mathcal{Q}^{(\nabla\si)^2}_{\mathsf{ab}}(k_s;\tau_1,\tau_2)=&\ \int\FR{\di^3\bm{q}}{(2\pi)^3}\Big[\big(\bm{q}\cdot(\bm{k}_s-\bm{q})\big)^2D^{\nn}_{\mathsf{ab}}(q;\tau_1,\tau_2)D^{\nn}_{\mathsf{ab}}(|\bm{k}_s-\bm{q}|;\tau_1,\tau_2)\n \\
    &+\partial_{\tau_1} \partial_{\tau_2}D^{\nn}_{\mathsf{ab}}(q;\tau_1,\tau_2)\partial_{\tau_1}\partial_{\tau_2}D^{\nn}_{\mathsf{ab}}(|\bm{k}_s-\bm{q}|;\tau_1,\tau_2)\Big](-\tau_1)^2(-\tau_2)^2.
\end{align} 
For directly-coupled vector boson loop generated by (\ref{eq_int_A}), we have
\begin{equation}\label{1L4PAAMoIn}
    \mathcal{Q}^{A^2}_{\mathsf{ab}}(k_s;\tau_1,\tau_2)=\int\FR{\di^3\bm{q}}{(2\pi)^3}\,g^{\mu\nu}g^{\mu'\nu'}D^A_{\mu\mu',\mathsf{ab}}(\bm q;\tau_1,\tau_2)D^A_{\nu\nu',\mathsf{ab}}(\bm{k}_s-\bm{q};\tau_1,\tau_2).
\end{equation}
Here $D_{\mu\mu',\aa\bb}^{A}$ are bulk propagators of the massive vector boson $A_\mu$. The explicit form of these propagators  can be found in \cite{Frob:2013qsa} but we never need them in this work. 

In above examples, the field running in the bubble can have zero or nonzero spins, but all  composite operators $\mathcal{O}$ happen to be scalars. Nevertheless, the spectral decomposition can be performed for operators with nontrivial tensor structure \cite{Loparco:2023rug}. This tensorial extension allows us to compute bubble loop diagrams with essentially arbitrary noncovariant couplings. For instance, we can consider the 4-point correlator from the following interaction:
\bge
\label{eq_LagPhi2SigmaPrime2}
  \Delta\mathcal{L}=-\FR14 a^2\phi_c^2\si'^2.
\ede
Although the operator $\si'^2$ by itself is not dS-covariant, it can be identified as the 00-component of a covariant tensor operator $\mathcal{O}_{\mu\nu}=(\nabla_\mu\si)(\nabla_\nu\si)$. Using this tensor embedding, we can in principle compute bubble graphs with arbitrary boost-breaking interactions. We will not pursue this direction in this work, but we think that this is an important point to make. We leave a more systematic study along this direction for a future work.

A direct computation of 1-loop correlators is challenging. We use the \emph{spectral decomposition} to circumvent the difficulty. 
The central idea of spectral decomposition is best explained in position space, where the 1-loop momentum integral is Fourier transformed to a 2-point function and is essentially a product of two position-space propagators. Then, spectral decomposition lets us rewrite this 2-point function as a linear combination (spectral integral) of scalar propagators $D^\si_{\aa\bb}$ of different masses: 
\begin{equation}\label{eq_SpeDec_Prop}
  \FR{1}{2}\la\mc O(X_1)\mc O(X_2)\ra_{\aa\bb,\rm{conn}}=\int_{-\infty}^{\infty}\di\nn'\rho^\mc{O}(\nn')D_{\aa\bb}^{\nn'}(X_1,X_2).
\end{equation}
Here, a factor of $1/2$ is introduced in line with the literature \cite{Xianyu:2022jwk,Zhang:2025nzd}, and ``conn'' means the connected part. Also, we are schematically writing a position-space propagator as $D_{\aa\bb,\mu_1\cdots\mu_j}$ where the Lorentz indices $(\mu_1,\cdots\mu_j)$ may be from either indices of spinning fields (such as in $A_\mu$) or spatial derivatives (such as in $\pd_\mu\si$). The three examples (\ref{eq_int_si}), (\ref{eq_int_patsi}), and (\ref{eq_int_A}) considered in this work are all scalar couplings, where the bubble loop mediates no total angular momentum. This corresponds to the cases where all Lorentz indices fully contracted in (\ref{eq_SpeDec_Prop}).\footnote{Generalization to bubbles carrying nonzero total angular momentum is straightforward: One simply replace the spectral functions and the propagator basis to accommodate this fact. See, e.g., \cite{Loparco:2023rug} for details.} The spectral function $\rho^{\mathcal{O}}$ represents the weight of $D_{\aa\bb}^{\nn'}$ in the spectral decomposition. If these spectral functions are known (more later), we can Fourier-transform the 2-point function back to momentum space, where the momentum integral is converted to the spectral integral:
\begin{equation}\label{eq_SpeInt_Prop}
    \mathcal{Q}^{\mathcal{O}}_{\mathsf{ab}}(k_s;\tau_1,\tau_2)=\int_{-\infty}^\infty\di\wt{\nu}'\rho^{\mathcal{O}}(\wt{\nu}')D^{\nn'}_{\mathsf{ab}}(k_s;\tau_1,\tau_2).
\end{equation}
Substituting (\ref{eq_SpeInt_Prop}) into the loop correlator (\ref{eq_LoopSeedDef}) and comparing it with the tree-level correlator (\ref{eq_TreeSeedDef}), we immediately find the bubble correlators of interest can be written as a superposition of tree-level correlators:\footnote{Here we are assuming that all integrals are either well convergent or properly regularized so that the orders of integrals can be exchanged freely.}
\begin{equation}\label{eq_SpeInt_Corre}
  \mathcal{J}^{\mathcal{O}}(r_1,x)=\FR{1}{2}\int_{-\infty}^\infty\mathrm{d}\wt{\nu}'\rho^{\mathcal{O}}(\wt{\nu}')\,\mathcal I^{\wt\nu'}(r_1,x).
\end{equation}
Again, we label the tree-level correlator with a superscript $\nn'$ to show its dependence on the integration variable. Since the tree-level correlators have been bootstrapped as reviewed in the previous section, we can try to finish the spectral integral, provided we know the spectral functions $\rho^{\mathcal{O}}$. 

Therefore, a central ingredient we get from the spectral decomposition is the spectral function $\rho^{\mathcal{O}}$. \cite{Loparco:2023rug} provides a method which allows us to compute the spectral function in the embedding space of dS. For the 1-loop processes studied in this work, we review the computation of required spectral functions $\rho^{\mathcal{O}}$ in App.\;\ref{appd_spectral} and collect the results here: 
\begin{align}
   \label{SpecSigma2}
    \rho^{\sigma^2}(\wt{\nu}')=&~\FR{\wt{\nu}'\sinh(\pi\wt{\nu}')}{16\pi^5\Gamma\big(\fr32+\ii\wt{\nu}'\big)\Gamma\big(\fr32-\ii\wt{\nu}'\big)}\prod_{\aa,\bb,\cc=\pm}\Gamma\Big[ \fr34+\fr12\ii(\aa+\bb)\wt{\nu} +\fr12\ii\cc\wt{\nu}' \Big], \\
    \label{SpecDeriv}
    \rho^{(\nabla\si)^2}(\wt{\nu}')=&~\FR{\wt{\nu}'(\fr94+2\wt{\nu}^2-\wt{\nu}'^2 )^2\sinh(\pi\wt{\nu}')}{64\pi^5\Gamma\big(\fr32+\ii\wt{\nu}'\big)\Gamma\big(\fr32-\ii\wt{\nu}'\big)} \prod_{\aa,\bb,\cc=\pm}\Gamma\Big[ \fr34+\fr12\ii(\aa+\bb)\wt{\nu} +\fr12\ii\cc\wt{\nu}' \Big], \\
    \label{SpecA2}
    \rho^{A^2}(\wt{\nu}')=&~ \FR{\wt{\nu}' (297+120\wt{\nu}'^2+16\wt{\nu}'^4+336\wt{\nu}_A^2-64\wt{\nu}'^2\wt{\nu}_A^2+192\wt{\nu}_A^4 )\sinh(\pi\wt{\nu}')}{1024\pi^5\big(\fr12+\ii\wt{\nu}_A\big)^2\big(\fr12-\ii\wt{\nu}_A\big)^2\Gamma\big(\fr32+\ii\wt{\nu}'\big)\Gamma\big(\fr32-\ii\wt{\nu}'\big)}\n\\
    &\times\prod_{\aa,\bb,\cc=\pm}\Gamma\Big[ \fr34+\fr12\ii(\aa+\bb)\wt{\nu}_A +\fr12\ii\cc\wt{\nu}' \Big].
\end{align}
Here, $\nn=\sqrt{m^2_\si-9/4}$ is the mass parameter of the exchanged massive scalar $\si$, and $\nn_A=\sqrt{m^2_A-1/4}$ is the mass parameter of the exchanged massive vector $A_\mu$. 

From the explicit expressions, it is clear that the spectral integrand is a meromorphic function of the spectral parameter $\wt\nu'$, implying that we can try to perform the spectral integral (\ref{eq_SpeInt_Corre}) by deforming the contour and collecting residues of appropriate poles.

Since the original contour is along the whole real axis, we can choose to close the contour from either the upper or lower plane of $\nn'$, depending on the convergence of the integrand, then collect all relevant poles.
The spectral functions introduced above have a common structure as products of $\Gamma$ functions, and pole structures of them are similar. All relevant poles of $\rho^{\mathcal{O}}$ can be classified as below.
\begin{align}
  &\text{Poles} && \text{lower\ }\wt{\nu}'\text{\ plane} && \text{upper\ }\wt{\nu}'\text{\ plane} \n \\
  &\label{eq_pole_A}\text{Set A:} &&\wt{\nu}'=-3\ii/2-2\nn-2\ii n  &&\wt{\nu}'=3\ii/2+2\wt{\nu}+2\ii n, \\
  &\label{eq_pole_B}\text{Set B:} &&\wt{\nu}'=-3\ii/2+2\wt{\nu}-2\ii n  &&\wt{\nu}'=3\ii/2-2\wt{\nu}+2\ii n,\\
  &\label{eq_pole_C}\text{Set C:} &&\wt{\nu}'=-3\ii/2-2\ii n  &&\wt{\nu}'=3\ii/2+2\ii n,
\end{align}
where $n=0,1,2,\cdots$.
Note that for $\rho^{A^2}$ we need to replace all $\nn$ in Set A and Set B with $\nn_A$.

Also, the pole structure of tree-level correlator can be read off from results in Sec.\;\ref{sec_dispersion_tree}, which are listed below\footnote{If we separate the tree correlator into nonlocal signal, local signal, and the background, then, the local signal and the background respectively possess poles at $\nn'=\pm\ii(1/2+n)$. However, the residues from these two parts cancel each other at every pole, showing that $\nn'=\pm\ii(1/2+n)$ are not poles of the entire tree correlator. A detailed analysis can be found in App.\;B of \cite{Zhang:2025nzd}.}.
\begin{align}
  &\text{Poles} && \text{lower\ }\wt{\nu}'\text{\ plane} && \text{upper\ }\wt{\nu}'\text{\ plane} \n \\
  &\label{eq_pole_D}\text{Set D}&&\nn'=-\ii n&&\nn'=\ii n.
\end{align}
However, this set of poles coincide precisely with the zeros of the spectral function due to a common factor $\sinh(\pi\nn')$ in $\rho^{\mathcal{O}}$  and thus contribute nothing to the final result.

Before performing any explicit calculation, we can make a key observation that different sets of poles listed above contribute components with different analytical structures in the 1-loop correlator. In particular, poles of Set A in (\ref{eq_pole_A}) and  Set B in (\ref{eq_pole_B}) have real parts $\text{Re}\,\nn'=\pm2\nn$. So, when collecting residues from these poles, the tree-level nonlocal signal (which contains nonanalytic factor $r_1^{\pm\ii\nn'}$) leads to nonanalyticities as $r_1^{\pm2\ii\nn}$ in the loop correlator, and thus belongs to the nonlocal signal of the bubble correlator. Performing similar analyses for all these poles, we find that the entire 1-loop nonlocal signal is contributed  by retaining only the tree-level nonlocal signal and collecting residues of poles only from Sets A in (\ref{eq_pole_A}) and B in (\ref{eq_pole_B}): 
\begin{align}
  \mathcal{J}^{\mathcal{O}}_{\text{NS}}(r_1,x)=\pm2\pi\ii\sum_{\text{Set AB}}\mathop{\text{Res}}\Big[\FR{1}{2}\rho^{\mathcal{O}}(\nn')\,\mathcal{I}^{\nn'}_{\text{NS}}(r_1,x)\Big]  .
\end{align} 
The plus or minus sign here depends on the orientation of the integration contour: The positive (negative) sign is adopted when closing the contour in the upper (lower) $\nn'$ plane. Using the expression for $\mc{I}_\text{NS}^{\wt\nu'}$ in (\ref{eq_TreeNonlocal}) to rewrite it as a sum of nonlocal signals of two modes with fixed scaling dimensions, we get:
\begin{align}\label{eq_SpeInt_NS}
  \mc J^{\mc O}_{\text{NS}}(r_1,x)=&~\pi\ii \sum_{\text{Set AB, upper}}\mathop{\rm{Res}}\Big[ \rho^{\mc O}(\nn')\mb I^{3/2-\ii\nn'}_{\text{NS}}(r_1,x)\Big]\n\\
  &~-\pi\ii\sum_{\text{Set AB, lower}}\mathop{\rm{Res}}\Big[ \rho^{\mc O}(\nn')\mb I^{3/2+\ii\nn'}_{\text{NS}}(r_1,x)\Big].
\end{align}
In this way, we turn the 1-loop nonlocal signal $\mathcal{J}^{\mathcal{O}}_{\text{NS}}$ into the following sum:
\begin{keyeqn}
\begin{align}\label{eq_NS_spectral_sum}
  \mathcal{J}^{\mathcal{O}}_{\text{NS}}(r_1,x)=\sum_{\Delta=\Delta_\pm}\sum_{n=0}^{\infty}\mathop{\mathcal{C}^{\mathcal{O}}_{\Delta,n}}\mathbf{I}^{2\Delta+2n}_{\text{NS}}(r_1,x),
\end{align}
\end{keyeqn}
That is, the 1-loop nonlocal signal can be expressed as a discrete sum of nonlocal signals of tree graphs mediating quasinormal modes of scaling dimensions $2\Delta_\pm+2n=3+2n\pm2\ii\wt\nu$, weighted by the \emph{spectral coefficients} $\mc{C}_{\Delta,n}^{\mathcal{O}}$, which is computed from the density function by: 
\begin{align}\label{eq_coefficient}
  \mc C^{\mc O}_{\Delta,n}=\mathop{\rm{Res}}_{\nn'=2\ii\Delta+2\ii n}\Big[\pi\ii\times\rho^{\mc O}(\nn')\Big]-\mathop{\rm{Res}}_{\nn'=-2\ii\Delta-2\ii n}\Big[\pi\ii\times\rho^{\mc O}(\nn')\Big],
\end{align}
where $\Delta=\Delta_\pm=3/2\pm\ii\nn$ (or $3/2\pm\ii\nn_A$ for the vector case) and $n=0,1,2,\cdots$. 

In principle, we can go on and consider other sets of poles listed above and collect all residues properly to finish the computation of $\mathcal{J}^{\mathcal{O}}$. However, the computation is significantly more involved and the results are not as simple and illuminative as $\mathcal{J}^{\mathcal{O}}_{\text{NS}}$. The simplicity of $\mathcal{J}^{\mathcal{O}}_{\text{NS}}$ as a discrete sum is the major technical reason for us to compute the full loop integral $\mathcal{J}^{\mathcal{O}}$ from its nonlocal signal $\mathcal{J}^{\mathcal{O}}_{\text{NS}}$ with a dispersive approach. Thus, the discrete sum (\ref{eq_NS_spectral_sum}) will be an important bootstrap ingredient for our spectral dispersion program, to be discussed in the next section. 

\section{Spectral Dispersion as a Bootstrap Strategy}\label{sec_dispersion_method}

With the dispersion relation and spectral decomposition introduced in the last two sections, now we are ready to discuss the spectral dispersion, which is a combination of the two techniques. 

Our strategy is based on the peculiar singularity structure of inflation correlators which is very different from flat-space scattering amplitudes, as reviewed in Introduction. For inflation correlators, both tree and loop graphs generally contain branch points. In particular, for 4-point correlators, the single-exchange tree graph and the 1-loop bubble graph (and also multi-loop banana graphs) generally share identical branch points. This, in particular, includes the nonlocal signal branch point which plays a central role in the line dispersion relation.

Through studies in recent years, it is now clear that the nonlocal signal in an inflation correlator is generated from intermediate degrees going on-shell, so it is very similar to the on-shell pole in flat-space scattering amplitudes. In particular, it satisfies an on-shell factorization theorem to all loop orders, which shows that the nonlocal signal is generally much simpler than the whole result. From this viewpoint, the line dispersion relation is a bootstrap strategy based on the purely on-shell data together with the analyticity of the whole correlator and thus is very similar in spirit to other well-known on-shell techniques such as BCFW and generalized unitarity cut. 

Given that the bubble loop graphs share the identical singularity structure with tree graphs, it is tempting to apply the line dispersion relation to bubble loop graphs, too: We start from the on-shell data of a bubble loop graph, namely its nonlocal signal, computed from the spectral decomposition (\ref{eq_NS_spectral_sum}). Then, we compute the whole loop correlator by finishing a dispersion integral similar to (\ref{eq_QNMLineDispersion}). 

Indeed, the above strategy has been applied in \cite{Liu:2024xyi} to bootstrap a 3-point function mediated by a massive bubble loop. We can well apply it to 4-point bubble loop correlators, which is logically trivial but technically involved. Thus, we seek to avoid this technical complication by making a further observation: The nonlocal signal of a bubble loop, as shown in (\ref{eq_NS_spectral_sum}), can be expressed as a discrete sum over nonlocal signals of quasinormal mode. Then, for each quasinormal mode, the whole tree correlator is fully determined by its nonlocal signal alone, showing a one-to-one relation between $\mb{I}_\text{NS}^{\Delta}$ and $\mb{I}^{\Delta}$. This motivates us to guess that we may simply remove the ``NS'' label on both sides of (\ref{eq_NS_spectral_sum}), so that the whole loop graph is also a discrete sum over tree graphs mediating quasinormal modes, with the \emph{same} spectral coefficients:
\begin{align}
\label{eq_SpecDispersionAnsatz}
  \mathcal{J}^{\mathcal{O}} (r_1,x)\stackrel{?}{=}\sum_{\Delta=\Delta_\pm}\sum_{n=0}^{\infty}\mathop{\mathcal{C}^{\mathcal{O}}_{\Delta,n}}\mathbf{I}^{2\Delta+2n} (r_1,x).
\end{align}
Below, we will show that this guess is almost correct, and the incorrect part is only a finite number of local terms whose coefficients must be determined by a renormalization condition. 

To derive the correct version of (\ref{eq_SpecDispersionAnsatz}), we begin with the spectral decomposition (\ref{eq_SpeInt_Corre}) together with the analytical structure of tree correlators (\ref{eq_TreeDiscRelation}). From these two equations, we can read off the discontinuity of the 1-loop correlator $\mc{J}^{\mc{O}}$ similar to the tree graphs:
\begin{align}
\label{eq_DiscJ}
  \mathop{\text{Disc}}\limits_{r_1}\mathcal{J}^{\mathcal{O}}(r_1,x)&=\mathop{\text{Disc}}\limits_{r_1}\bigg[\FR{1}{2}\int_{-\infty}^{\infty}\di\nn'\rho^{\mathcal{O}}(\nn')\,\mathcal{I}^{\nn'}(r_1,x)\bigg]\n\\
  &=\FR{1}{2}\int_{-\infty}^{\infty}\di\nn'\rho^{\mathcal{O}}(\nn')\mathop{\text{Disc}}\limits_{r_1}\,\mathcal{I}^{\nn'}(r_1,x)\n\\
  &=\FR{1}{2}\int_{-\infty}^{\infty}\di\nn'\rho^{\mathcal{O}}(\nn')\mathop{\text{Disc}}\limits_{r_1}\Big[\mathcal{I}^{\nn'}_{\text{NS}}(r_1,x)-\mathcal{I}^{\nn'}_{\text{NS}}(-r_1,x)\Big]\n\\
  &=\mathop{\text{Disc}}\limits_{r_1}\bigg\{\FR{1}{2}\int_{-\infty}^{\infty}\di\nn'\rho^{\mathcal{O}}(\nn')\Big[\mathcal{I}^{\nn'}_{\text{NS}}(r_1,x)-\mathcal{I}^{\nn'}_{\text{NS}}(-r_1,x)\Big]\bigg\}\n\\
  &=\mathop{\text{Disc}}\limits_{r_1}\Big[\mathcal{J}_{\text{NS}}^{\mathcal{O}}(r_1,x)-\mathcal{J}^{\mathcal{O}}_{\text{NS}}(-r_1,x)\Big].
\end{align} 
Here we interchange freely the order of taking discontinuity and the spectral integral, assuming that the spectral integral itself is properly regulated and everything is finite. 

Then, with (\ref{eq_DiscJ}) at hand, we can process in the same way as Sec.\;\ref{sec_dispersion_tree} and build a ``line'' dispersion integral which recovers the full correlator from its on-shell data, namely, the nonlocal signal:
\begin{align}\label{eq_dispersion_bubble}
  \mathcal{J}^{\mathcal{O}}(r_1)=\int_{-\infty}^{0}\FR{\di r}{2\pi\ii}\FR{\displaystyle\mathop{\text{Disc}}\big[\mathcal{J}_{\text{NS}}^{\mathcal{O}}(r)-\mathcal{J}_{\text{NS}}^{\mathcal{O}}(-r)\big]}{r-r_1}\prod_{i=1}^N\FR{r_1-a_i}{r-a_i}.
\end{align}
Here we have introduced $N$ subtractions at $r=a_1,\cdots,a_n$ where $N$ should be chosen such that the integrand decreases fast enough at infinity, with which the dispersion integral will give us a result free of UV divergence. We introduce this additional subtraction because we do not a priori know the UV behavior of the loop correlator $\mathcal{J}^{\mathcal{O}}$ without specifying $\mathcal{O}$. However, we do know that, for any well-defined local operator $\mathcal{O}$, computing $\mathcal{J}^{\mathcal{O}}$ at 1-loop order (or rather, any finite loop order) only generate power-law divergences in the UV, which corresponds to a power-law divergence in the integral over the large circle in Fig.\;\ref{fig_dispersion}. Therefore, $N$ subtraction with finite integer $N$ is guaranteed to remove any UV divergence from the large circle.

Next, we compute (\ref{eq_dispersion_bubble}) by substituting (\ref{eq_NS_spectral_sum}) into the integrand and exchange the order of integration and summations:
\begin{align}\label{eq_dispersion_QNmode}
   \mathcal{J}^{\mathcal{O}}(r_1)= 
   \sum_{\Delta=\Delta_{\pm}}\sum_{n=0}^{\infty} \mathop{\mathcal{C}^{\mathcal{O}}_{\Delta,n}}\int_{-\infty}^{0}\FR{\di r}{2\pi\ii(r-r_1)}\text{Disc} \Big[\mathbf{I}_{\text{NS}}^{2\Delta+2n}(r)-\mathbf{I}_{\text{NS}}^{2\Delta+2n}(-r)\Big]   \prod_{i=1}^N\FR{r_1-a_i}{r-a_i}.
\end{align}
After the change of integration and summations, the integral is nothing but a dispersion integral of tree-level quasinormal mode, with an undetermined subtraction scheme. Ignoring the issue of subtraction, the result of this integral would simply be the whole result of the tree graph mediating a quasinormal mode. If this is true, then our guess (\ref{eq_SpecDispersionAnsatz}) is justified.

However, it is not entirely correct to ignore the issue of subtraction. The easiest way to see the issue is the following: The tree correlator $\mb{I}^{\Delta}$ in (\ref{eq_SpecDispersionAnsatz}) is itself computed from the dispersion integral (\ref{eq_QNMLineDispersion}), which contains a single subtraction. So, if we sum over these tree correlators as in (\ref{eq_SpecDispersionAnsatz}), the result would be equivalent to taking single subtraction in (\ref{eq_dispersion_QNmode}) as well. We know this cannot be always correct, since the dispersion integral for the loop correlator $\mathcal{J}^{\mathcal{O}}$ could be more divergent than its tree-level counterpart in the UV and thus may require more subtractions. Therefore, our naive expectation in (\ref{eq_SpecDispersionAnsatz}) has a risk of insufficient subtractions. 

Insufficient subtractions mean that the dispersion integral is divergent in the UV. In (\ref{eq_SpecDispersionAnsatz}), this divergence must be shifted to the divergence of $n$-summation which renders the whole expression meaningless. On the other hand, we also know that these UV divergences must be absent when the loop correlators are properly renormalized. In particular, at the 1-loop order, the divergence must be removed by a finite number of counterterms. This suggests that we can simply use (\ref{eq_SpecDispersionAnsatz}) to compute the loop correlators. The result would probably contain divergent $n$-summation. It would be important to identify the divergent pieces in this sum and to show that they have identical kinematic dependence with contact graphs produced by correct local counterterms. Once this is achieved, we can simply throw away the divergent pieces, and supplement the result with a finite number of local contact graphs with coefficients determined by a renormalization condition. 

With the above consideration in mind, we can now rewrite our naive guess (\ref{eq_SpecDispersionAnsatz}) in a correct manner:
\begin{keyeqn}
\begin{align}\label{eq_bubble_spectral_sum}
  &\mathcal{J}^{\mathcal{O}}(r_1,x)=\sum_{\Delta=\Delta_{\pm}}\sum_{n=0}^{\infty}\mathop{\mathcal{C}^{\mathcal{O}}_{\Delta,n}}\mathbf{I}^{2\Delta+2n}(r_1,x)+\text{local terms},
\end{align}
\end{keyeqn}
where the local terms mean the analytic functions produced by local counterterms. This is the key formula for our spectral dispersion. In the next two sections, we will use this formula to bootstrap a few 1-loop bubble graphs at 3-point and 4-point levels. Along the way, it is important to make sure that we can identify the divergent pieces in the $n$-sum and show that they can be subtracted away by correct counterterms. We will see that this can indeed be achieved.

\section{An Example of Three-Point Bubble Graph}\label{sec_dispersion_method_3pt}

\begin{figure}[t]
  \centering 
  \includegraphics[width=0.99\textwidth]{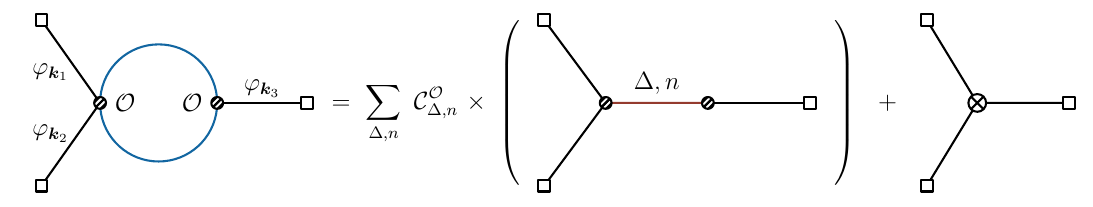} 
  \caption{The spectral dispersion of the 3-point correlator of the massless scalar $\varphi$ with 1-loop bubble exchange.
  }
\label{fd_3pt_bubble}
\end{figure}

In this section, we illustrate our bootstrap procedure with an example of 3-point bubble graph with a pair of directly coupled massive scalars, as shown in Fig.\;\ref{fd_3pt_bubble}. This example has been worked out using both the spectral \cite{Xianyu:2022jwk} and dispersive techniques \cite{Liu:2024xyi}. We use it to show explicitly the subtleties discussed previously and as a verification of our general strategy.

For simplicity, we assume that the two vertices in Fig.\;\ref{fd_3pt_bubble} are respectively from the two terms in the following Lagrangian for a massless scalar $\varphi$ and a massive scalar $\sigma$: 
\begin{align}\label{eq_int_3pt_si}
  \Delta\mathcal{L}=-\FR{1}{4}a^2\varphi'^2\sigma^2-\FR{1}{2}a^3\varphi'\sigma^2.
\end{align}
This interaction can give rise to a 1-loop bubble diagrams contributing to the 3-point correlator of $\varphi$, shown in Fig.\;\ref{fd_3pt_bubble}. 
Similar to what we did before, we can write this 3-point correlator in terms of a dimensionless integral $\mathcal{K}$:
\begin{align}
  \la\varphi_{\bm{k}_1}\varphi_{\bm{k}_2}\varphi_{\bm{k}_3}\ra_{\si^2}'=\FR{1}{8k_1k_2k_3^4}\mathcal{K} \Big(\FR{2k_{3}}{k_{123}}\Big),
\end{align}
where the integral $\mathcal{K}$ is defined as
\begin{align}\label{eq_K3ptO}
    \mathcal{K}^{\mathcal{O}}_{\text{3-pt}}\Big( \FR{2k_3}{k_{123}} \Big)\equiv \FR{k_3^5}{2}\!\sum_{\aa,\bb=\pm}(-\aa\bb)\int_{-\infty}^0\FR{\di\tau_1 \di\tau_2}{(-\tau_2)^2}\,\mathcal{Q}^{\mathcal{O}}_{\aa\bb}\big(k_3;\tau_1,\tau_2\big)e^{\ii\aa k_{12}\tau_1+\ii \bb k_{3}\tau_2},
\end{align}
and for brevity we write $\mc K\equiv\mc K^{\si^2}_{\text{3-pt}}$. Since there are only two independent momentum variables $k_{12}$ and $k_3$ on the right-hand side, the dimensionless integral $\mathcal{K}$ must be dependent only on one momentum ratio, which we can choose to be $u\equiv2k_3/k_{123}$. If we think of this 3-point function as a folded limit, $k_4\to 0$, of a 4-point bubble (c.f.\ Fig.\;\ref{fd_4pt_bubble} albeit with different external fields), then $k_s=|\bm k_3+\bm k_4|\to k_3$ in this limit. Consequently, the nonlocal and local signals are indistinguishable and thus combined into a single ``signal.'' Similar to the 4-point nonlocal signal, the 3-point signal part has a neat form:
\begin{align}\label{eq_bubble_3pt_signal}
  \mathcal{K}_\text{S}(u)=\sum_{\Delta=\Delta_{\pm}}\sum_{n=0}^{\infty}\mathop{\mathcal{C}^{\si^2}_{\Delta,n}}\mathbf{I}^{2\Delta+2n}_{\text{3-pt,\,S}}(u),
\end{align}
where the spectral coefficient $\mc C^{\mc O}_{\Delta,n}$ can be obtained using \eqref{eq_coefficient} and the result is:
\begin{align}
\label{eq_Csigma2}
  \mathcal{C}_{\Delta,n}^{\sigma^2}&=\mathop{\rm{Res}}_{\nn'=2\ii\Delta+2\ii n}\Big[\pi\ii\times\rho^{\si^2}(\nn')\Big]-\mathop{\rm{Res}}_{\nn'=-2\ii\Delta-2\ii n}\Big[\pi\ii\times\rho^{\si^2}(\nn')\Big]\n\\
  &=\FR{(4\Delta+4n-3)(\tan(\pi\Delta)^2-1)\,(n+1)_{1/2}\,(2\Delta+n-2)_{1/2}}{32\pi^2(\Delta+n)_{1/2}\,(\Delta+n-1)_{1/2}},
\end{align} 
and $\mathbf{I}_{\text{3-pt,\,S}}^{\Delta}$ is the 3-point tree-level quasinormal mode of signal,\footnote{This result has been presented in \cite{Qin:2023ejc} where the signal part of the 3-point function is contained in $\wt{\mc I}^{0,-2}(u,1)$ and is written as $\mb I_{\text{3-pt,\,S}}^{\Delta_+}+\mb I_{\text{3-pt,\,S}}^{\Delta_-}$.}
\begin{align}\label{eq_tree_3ptS_QNMode}
  \mathbf{I}^{\Delta}_{\text{3-pt,\,S}}(u)=-\FR{\pi}{4}\FR{\cos(\pi\Delta)+1}{\sin(2\pi\Delta)}\,u^{\Delta+1}\,{}_2\mathcal{F}_1\bigg[\bgm\Delta+1,\Delta-1\\2\Delta-2\edm\bigg|u\bigg],
\end{align} 

At this point, we can proceed in two different ways as discussed before.

\emph{Option 1:} We can develop a dispersion integral directly for the 3-point bubble integral $\mathcal{K}$. This was the approach adopted in 
\cite{Liu:2024xyi} where it was shown that the integral $\mathcal{K}$ has a single branch cut on the entire negative real axis of $u$, and the discontinuity of this cut is contributed by the signal alone.\footnote{The integral $\mathcal{K}$ here was called $\mathcal{J}$ in \cite{Liu:2024xyi}.} Therefore, with the signal $\mathcal{K}_\text{S}$ in (\ref{eq_bubble_3pt_signal}) known, we can recover $\mathcal{K}$ itself from $\text{Disc}\,\mathcal{K}_\text{S}$ via:
\begin{align}\label{eq_dispersion_bubble_3pt}
  \mathcal{K}(u)=\int_{-\infty}^{0}\FR{\di u'}{2\pi\ii}  \FR{\mathop{\text{Disc}}\mathcal{K}_\text{S}(u') }{(u'-u)}\prod_{i=1}^N\FR{u-a_i}{u'-a_i}.
\end{align}
Again, we have introduced $N$ subtractions to make the integral convergent at infinity. It turns out that this integral is simple enough to be done directly. In fact, from (\ref{eq_bubble_3pt_signal}) it is easy to see that the signal $\mathcal{K}_\text{S}\sim u^{4\pm 2\ii\wt\nu}$ as $u\to 0$, so that we can utilize this fourth-order zero to make a fourth subtraction, namely, with $N=4$ and all $a_i=0$ in (\ref{eq_dispersion_bubble_3pt}), so that the dispersion integral simplifies to:
\begin{align}\label{eq_dispersion_bubble_3pt_u^4}
  \mathcal{K} (u)=\int_{-\infty}^{0}\FR{\di u'}{2\pi\ii}\FR{\displaystyle\mathop{\text{Disc}}\mathcal{K} (u')}{(u'-u) }\Big(\FR{u}{u'}\Big)^4.
\end{align}
Finishing this integral as detailed in \cite{Liu:2024xyi}, we get the full answer as $\mathcal{K}=\mathcal{K}_\text{S}+\mathcal{K}_\text{BG1}$, where the signal $\mathcal{K}_\text{S}$ is simply (\ref{eq_bubble_3pt_signal}), and the background $\mathcal{K}_\text{BG1}$ reads 
\begin{align}
\label{eq_KBG1}
  \mathcal{K}_{\text{BG}1}(u)=\sum_{\Delta=\Delta_{\pm}}\sum_{n=0}^{\infty}   \FR{\pi\cot(\pi\Delta)}{8\cos(2\pi\Delta)}\mathcal{C}^{\si^2}_{\Delta,n}u^4\times{}_3\mathcal{F}_2\bigg[\bgm1,2,4\\1+2\Delta+2n,4-2\Delta-2n\edm\bigg|u\bigg].
\end{align}
The $n$-summation is well convergent for $u$ in the physical region. There is no UV divergence here thanks to the fourth-order subtraction. 

\emph{Option 2:} Alternatively, as discussed previously, with (\ref{eq_bubble_3pt_signal}) known, we can get the background $\mathcal{K}_\text{BG2}$ (BG2 meaning background from Option 2) by summing over the background for each quasinormal mode. This amounts to replacing 
$\mathbf{I}^{2\Delta+2n}_{\text{3-pt,\,S}}(u)$ by $\mathbf{I}^{2\Delta+2n}_{\text{3-pt,\,BG}}(u)$ in (\ref{eq_bubble_3pt_signal}):
\begin{align}\label{eq_bubble_3pt_BG2}
  \mathcal{K}_\text{BG2}(u)=\sum_{\Delta=\Delta_{\pm}}\sum_{n=0}^{\infty}\mathop{\mathcal{C}^{\si^2}_{\Delta,n}}\mathbf{I}^{2\Delta+2n}_{\text{3-pt,\,BG}}(u),
\end{align}
and $\mathbf{I}^{\Delta}_{\text{3-pt,\,BG}}(u)$ can be obtained from $\mathbf{I}^{\Delta}_\text{BG}(r,x)$ in (\ref{eq_tree_BG_QNmode}) by taking $x=r$ and using an identity of ${}_2\text{F}_1$ to change the variable from $r$ to $u$. The result is  
\begin{align}
\label{eq_KBG2}
  \mathcal{K}_\text{BG2}(u)=\sum_{\Delta=\Delta_{\pm}}\sum_{n=0}^{\infty} \FR{\pi\cot(\pi\Delta)}{8\cos(2\pi\Delta)}\mathop{\mathcal{C}^{\si^2}_{\Delta,n}} u^3\times{}_3\mathcal{F}_2\bigg[\bgm1,1,3\\2\Delta+2n,3-2\Delta-2n\edm\bigg|u\bigg].
\end{align}
Apparently, $\mathcal{K}_\text{BG2}\neq \mathcal{K}_\text{BG1}$. This is not surprising, since we should expect $\mathcal{K}_\text{BG2}=\mathcal{K}_\text{BG1}$ only up to analytic terms that cannot be determined by a dispersive computation alone. 

To compare $\mathcal{K}_\text{BG2}$ with $\mathcal{K}_\text{BG1}$ more carefully, we expand the hypergeometric factors ${}_3\mathcal{F}_2 [\cdots]$ in (\ref{eq_KBG1}) and (\ref{eq_KBG2}) as a Taylor series of $u$. Then, all terms starting from $u^4$ are the same in $\mathcal{K}_\text{BG1}$ and $\mathcal{K}_\text{BG2}$. The only disagreement is the $u^3$ term:
\begin{equation}
\label{eq_KBGdifference}
  \mathcal{K}_\text{BG2}(u)-\mathcal{K}_\text{BG1}(u)=\sum_{\Delta=\Delta_{\pm}}\sum_{n=0}^{\infty} \FR{\pi\cot(\pi\Delta)}{4\cos(2\pi\Delta)}\FR{\mathcal{C}^{\si^2}_{\Delta,n}}{\Gamma[2\Delta+2n,3-2\Delta-2n]} u^3. 
\end{equation}

The origin for this disagreement is clear: We need a single counterterm to renormalize the bubble loop graph,
\footnote{The UV divergence of any 1-loop diagram is not worsened by the presence of a finite spacetime curvature. The required number of counterterms to renormalize correlators at 1-loop order thus remains finite. This feature presumably persists to higher orders in loop expansion so far as the perturbation theory is concerned. That is, the correlators of interest are always renormalizable at a given loop order even for superficially nonrenormalizable models.}
\begin{align}
  \Delta\mathcal{L}\propto a \varphi'^3.
\end{align}
Computing the contact-graph contribution to the 3-point correlator from this interaction, we get exactly a result $\propto u^3$, showing that the coefficient of the $u^3$ should be determined by a renormalization condition rather than our calculation.

In fact, the $u^3$-term in $\mathcal{K}_\text{BG1}$ is 0 due to the fourth-order subtraction, while the $u^3$-term in $\mathcal{K}_\text{BG2}$ has a divergent $u$-sum, arising from a residual UV divergence, signaling insufficient subtractions.\footnote{More explicitly, for each qusinormal-mode contribution, the background $\mathbf{I}^{2\Delta+2n}_{\text{3-pt,\,BG}}(u)$ can be computed from the signal $\mathbf{I}^{2\Delta+2n}_{\text{3-pt,\,S}}(u)$ via a dispersion integral with third-order subtraction. So, if we naively take the sum of all backgrounds $\mathbf{I}^{2\Delta+2n}_{\text{3-pt,\,BG}}(u)$ weighted by the spectral coefficient $\mathcal{C}_{\Delta,n}^{\si^2}$, the result is equivalent to performing the integral (\ref{eq_dispersion_bubble_3pt}) with third-order subtraction with all $a_i=0$, which is clearly insufficient to remove the UV divergence from the large circle.} In any case, we should replace the coefficient of the $u^3$ term by an arbitrary and finite constant $C$ to be determined by renormalization condition, so that the final answer for our computation of the background should be written as:
\begin{align}
\label{eq_KBG}
  \mathcal{K}_{\text{BG}}(u)=C u^3+\sum_{\Delta=\Delta_{\pm}}\sum_{n=0}^{\infty} \FR{\pi\cot(\pi\Delta)}{8\cos(2\pi\Delta)}\mathcal{C}^{\si^2}_{\Delta,n} u^4\times{}_3\mathcal{F}_2\bigg[\bgm1,2,4\\1+2\Delta+2n,4-2\Delta-2n\edm\bigg|u\bigg].
\end{align}

To summarize, the example presented in this section shows that the two ways to do dispersive bootstrap yield the same result up to local terms to be determined by the renormalization condition. Since the second approach, summing over the quasinormal modes, is practically much simpler, we will adopt it in the next section to bootstrap more complicated examples. Along the way, we will encounter UV divergences in the form of divergent series. It is a crucial consistent check of our calculation that all these divergences must have identical kinematic structure as the graphs from local counterterms required to renormalize the graph.

\section{Bootstrapping 1-loop Bubble Correlators}\label{sec_dispersion_result}
 
In this section, we apply the spectral dispersion to bootstrap three 4-point 1-loop bubble correlators which are of main interest of this work. To reiterate our strategy, we first use the spectral decomposition to write the nonlocal signal of the bubble graphs as a discrete sum of nonlocal signals of quasinormal modes. From this procedure we will get the spectral coefficient $\mathcal{C}_{\Delta_\pm,n}^{\mathcal{O}}$. Then, we directly write down the result for the entire bubble loop as a linear superposition of full correlators mediated by quasinormal modes, weighted by $\mathcal{C}^{\mathcal{O}}_{\Delta_\pm,n}$. Finally, we identify local terms degenerate with counterterm contributions, which should be determined by renormalization condition. After removing these terms, the remaining pieces, including the nonlocal and local signal, and the irreducible background, are robust predictions of our calculation. We summarize this procedure in Fig.\;\ref{fd_4pt_bubble_dispersion}.

\subsection{Scalar bubble}

The first process we consider is the 4-point correlator with a 1-loop bubble massive scalar exchanged in the $s$-channel, where the massive scalar $\sigma$ is directly coupled to the conformal scalar $\phi_c$, coming from interaction (\ref{eq_int_si}). This graph has been worked out in \cite{Xianyu:2022jwk} based on purely spectral methods. Here, we will show that the use of spectral dispersion yields much simpler result. 

It is known that the dimensionless correlator $\mathcal{J}^{\mc{O}}(r_1,x)$ in (\ref{eq_LoopSeedDef}) can be separated into three pieces, similar to a tree-level graph with single exchange. Taking $\mathcal{O}=\si^2$, we have:
\begin{equation}
  \mathcal{J}^{\sigma^2}(r_1,x)=\mathcal{J}^{\sigma^2}_{\text{NS}}(r_1,x)+\mathcal{J}^{\sigma^2}_{\text{LS}}(r_1,x)+\mathcal{J}^{\sigma^2}_{\text{BG}}(r_1,x).
\end{equation}
Here we assume that the loop correlator is properly renormalized so that the background part $\mathcal{J}^{\sigma^2}_{\text{BG}}$ is finite and renormalization dependent. On the other hand, the nonlocal and local signals are both renormalization independent. 

To begin with, we compute the nonlocal signal of the loop graph. This is done by collecting  residues of poles contributing to the nonlocal signal in Sec.\;\ref{sec_spectral}, we find the nonlocal signal: 
\begin{align}
  \mathcal{J}^{\sigma^2}_{\text{NS}}(r_1,x)=\sum_{\Delta=\Delta_\pm}\sum_{n=0}^{\infty}\mathop{\mathcal{C}^{\sigma^2}_{\Delta,n}}\mathbf{I}_{\text{NS}}^{2\Delta+2n}(r_1,x),
\end{align}
where the spectral coefficient $\mathcal{C}_{\Delta,n}^{\sigma^2}$ has been computed before with the result given in (\ref{eq_Csigma2}). The nonlocal signal $\mathbf{I}^{\Delta}_{\text{NS}}$ for a single quasinormal mode of dimension $\Delta$ is given in (\ref{eq_tree_NS_QNmode}).

According to the discussion in Sec.\;\ref{sec_dispersion_method} and especially (\ref{eq_bubble_spectral_sum}), we immediately know that the full bubble loop correlator can be written as:
\begin{align}
\label{eq_Jsigma2unrenormalized}
  \mathcal{J}^{\sigma^2}(r_1,x)=\sum_{\Delta=\Delta_\pm}\sum_{n=0}^{\infty}\mathop{\mathcal{C}^{\sigma^2}_{\Delta,n}}\mathbf{I}^{2\Delta+2n}(r_1,x)+\text{local terms},
\end{align}
where the single-quasinormal-mode-exchange correlator $\mathbf{I}^{\Delta}=\mathbf{I}^{\Delta}_{\text{NS}}+\mathbf{I}^{\Delta}_{\text{LS}}+\mathbf{I}^{\Delta}_{\text{BG}}$ is the sum of (\ref{eq_tree_NS_QNmode}), (\ref{eq_tree_LS_QNmode}), and (\ref{eq_tree_BG_QNmode}). In particular, the $\mathbf{I}^{\Delta}_{\text{LS}}$ and $\mathbf{I}^{\Delta}_{\text{BG}}$ parts of the summand give rise to the local signal $\mathcal{J}_\text{LS}^{\sigma^2}(r_1,x)$ and the unregularized background (URBG) $\mathcal{J}_\text{URBG}^{\sigma^2}(r_1,x)$ of the bubble loop, respectively. 

It is straightforward to check that the nonlocal signal $\mathcal{J}_\text{NS}^{\sigma^2}$ and the local signal $\mathcal{J}_\text{LS}^{\sigma^2}$ are finite at physical region, while the background $\mathcal{J}_\text{URBG}^{\sigma^2}(r_1,x)$ is divergent for generic $r_1$ and $x$. Fortunately, it is easy to identify the divergent part from the background as a divergent sum in $n$. To this end, we expand the summand of (\ref{eq_Jsigma2unrenormalized}) in large $n$, so that the summation breaks into pieces such as $\sum_{n}1/n$, $\sum_{n} n$, $\cdots$. For (\ref{eq_Jsigma2unrenormalized}), we find that there is only one type of divergence, coming from the following sum in the large $n$ expansion of the summand: 
\begin{align}\label{eq_div_S}
  \text{Div}\Big[\mathcal{J}^{\sigma^2}_{\text{URBG}}\Big]\sim\sum_{n=1}^{\infty}\FR{1}{8\pi^2 n}\FR{r_1}{1+x}=\sum_{n=1}^{\infty}\FR{1}{8\pi^2 n}\FR{k_s}{k_{1234}}.
\end{align} 
It is reassuring that the shape of this divergent piece is identical to the contact graph generated from the following expected counterterm (again with coefficient neglected):
\begin{align}
  \Delta\mathcal{L}= a^4\phi_c^4.
\end{align}
So, the divergence would disappear after including the local contribution from this counterterm. 

Now, if we view the correlator as a series in $r_1$ with $x$ fixed, then the counterterm contributes only a $\order{r_1}$ term. 
Therefore, all higher order terms starting from $\order{r_1^3}$ are independent of renormalization condition. 
So, similar to the previous section, we can divide the background into two pieces, the renormalization-dependent $\order{r_1}$ piece $\mathcal{J}^{\sigma^2}_{\text{LBG}}$ which is the leading term in the squeezed limit $r_1\to 0$ and we call the \emph{leading background} (LBG), and the renormalization-independent piece $\mathcal{J}^{\sigma^2}_{\text{SLIBG}}$ starting from $\order{r_1^3}$ which we call the \emph{subleading irreducible background} (SLIBG):
\begin{equation}
  \mathcal{J}^{\sigma^2}_{\text{BG}}(r_1,x)=\mathcal{J}^{\sigma^2}_\text{LBG}(r_1,x)+\mathcal{J}^{\sigma^2}_{\text{SLIBG}}(r_1,x).
\end{equation}
Here, the leading background $\mathcal{J}^{\sigma^2}_\text{LBG}(r_1,x)$ is given by:
\begin{align}
\label{eq_Jsigma2LBG}
  &\mathcal{J}^{\sigma^2}_\text{LBG}(r_1,x)= 
   r_1 \sum_{\Delta=\Delta_{\pm}}\sum_{n,\ell=0}^{\infty}\bigg[\FR{\big(\sec(2\pi\Delta)+1\big)\mathcal{C}^{\si^2}_{\Delta,n}}{(\ell+2\Delta+2n-1)(\ell-2\Delta-2n+2)}-\FR{1}{16\pi^2 n}\bigg](-x)^\ell+\FR{Cr_1}{1+x}.
\end{align}
The summation here is well convergent for physical values of $x$, and we have included a contact-graph contribution with an arbitrary and finite coefficient $C$, which is determined by a renormalization condition. 

The subleading irreducible background $\mathcal{J}^{\sigma^2}_{\text{SLIBG}}$, on the other hand, is given by:  
\begin{align}
\label{eq_Jsigma2IBG}
  \mathcal{J}^{\sigma^2}_{\text{SLIBG}}(r_1,x)= 
  &\sum_{\Delta=\Delta_{\pm}}\sum_{n,\ell=0}^{\infty} \mathop{\mathcal{C}^{\sigma^2}_{\Delta,n}}\FR{\cos(\pi\Delta)^2}{8\cos(2\pi\Delta)}\FR{(\ell+1)_2}{(\fr{\ell}{2}+\Delta+n-\fr{1}{2})_2\,(\fr{\ell}{2}-\Delta-n+1)_2}\n\\
  &\times r_1^3\,(-x)^\ell\times{}_3\text{F}_2\bigg[\bgm1,\fr{\ell}{2}+\fr{3}{2},\fr{\ell}{2}+2\\\fr{\ell}{2}+\Delta+n+\fr{3}{2},\fr{\ell}{2}-\Delta-n+3\edm\bigg|r_1^2\bigg].
\end{align}
The irreducible background is independent of renormalization condition; Its form is completely fixed by the dispersion relation of the correlator and thus represents a robust prediction of our bootstrap program based on the on-shell data. The subleading irreducible background defined as above starts at $r_1^3$ as $r_1\to 0$ and is admittedly harder to measure than the leading background in realistic observations. 

It is curious to observe that there is also a piece of ``irreducible background'' in $\mathcal{J}^{\sigma^2}_\text{LBG}$ from the first term in (\ref{eq_Jsigma2LBG}): After finishing the sum over $n$ and $\ell$, the first term would be in a form of $f(x)r_1$ where $f(x)$ is in general a complicated function. However, we can always Laurent expand $f(x)$ at $x=-1$ to get a series $\sum\limits_{m}c_m (1+x)^m$. Then, all terms but $(1+x)^{-1}$ are also independent of renormalization, because they are of different forms from the local contribution $\propto r_1/(1+x)$ coming from the counterterm $\Delta\mathcal{L}= a^4\phi_c^4$. So, we can equally view $\sum\limits_{m\neq -1}c_m (1+x)^m$ as an ``irreducible background'' at the leading order in $r_1$, which would certainly be easier to detect than $\mathcal{J}^{\sigma^2}_\text{SLIBG}$. 

\subsection{Derivatively coupled scalar bubble}

Now, we consider a more complicated case: the conformal scalar correlator mediated by a pair of derivatively coupled massive scalars $\sigma$, which comes from the interaction in (\ref{eq_int_patsi}).
The spectral coefficient $\mathcal{C}^{(\nabla\sigma)^2}_{\Delta,n}$ for the operator $(\nabla\sigma)^2$ is certainly more complicated than $\mathcal{C}^{\si^2}_{\Delta,n}$ but is related to the latter in a simple way; The main complication in this case is that we need more counterterms to renormalize the corresponding 1-loop diagram, since $(\nabla\si)^2\phi_c^2$ is an operator of higher dimension than $\si^2\phi_c^2$. So we will have more independent local shapes coming from these counterterms, meaning that, in our under-subtracted bootstrap method, there will be more divergent sums which need to be identified in the calculation. 

Let the dimensionless loop correlator from $(\nabla\si)^2\phi_c^2$ operator be $\mathcal{J}^{(\nabla\sigma)^2}$. Again, it can be split into nonlocal signal, local signal, and the background:
\begin{equation}
  \mathcal{J}^{(\nabla\sigma)^2}=\mathcal{J}^{(\nabla\sigma)^2}_\text{NS}+\mathcal{J}^{(\nabla\sigma)^2}_\text{LS}+\mathcal{J}^{(\nabla\sigma)^2}_\text{BG}.
\end{equation}

We again start from the spectral decomposition in Sec.\;\ref{sec_spectral} and compute the nonlocal signal:
\begin{align}
  \mathcal{J}^{(\nabla\sigma)^2}_{\text{NS}}(r_1,x)=\sum_{\Delta=\Delta_{\pm}}\sum_{n=0}^{\infty}\mathop{\mathcal{C}^{(\nabla\sigma)^2}_{\Delta,n}}\mathbf{I}^{2\Delta+2n}_{\text{NS}}(r_1,x),
\end{align}
where the spectral coefficient $\mathcal{C}^{(\nabla\sigma)^2}_{\Delta,n}$ is again computed according to \eqref{eq_coefficient} and the result is 
\begin{align}
  \mathcal{C}^{(\nabla\sigma)^2}_{\Delta,n}=\big(\Delta^2+4n\Delta+(2n-3)n\big)^2\mathop{\mathcal{C}^{\sigma^2}_{\Delta,n}},
\end{align} 
and $\mathop{\mathcal{C}^{\sigma^2}_{\Delta,n}}$ is given in (\ref{eq_Csigma2}).

Then, the local signal $\mathcal{J}^{(\nabla\sigma)^2}_{\text{LS}}$ and the unregularized background $\mathcal{J}^{(\nabla\sigma)^2}_{\text{URBG}}$ are obtained as before:
\begin{align}
  \mathcal{J}^{(\nabla\sigma)^2}_{\text{LS}}(r_1,x)=&\sum_{\Delta_{\pm}}\sum_{n=0}^{\infty}\mathop{\mathcal{C}^{(\nabla\sigma)^2}_{\Delta_\pm,n}}\mathbf{I}^{2\Delta_\pm+2n}_{\text{LS}}(r_1,x),\\
  \label{eq_BG_DS_original}
  \mathcal{J}^{(\nabla\sigma)^2}_{\text{URBG}}(r_1,x)=&\sum_{\Delta_{\pm}}\sum_{n=0}^{\infty}\mathop{\mathcal{C}^{(\nabla\sigma)^2}_{\Delta_\pm,n}}\mathbf{I}^{2\Delta_\pm+2n}_{\text{BG}}(r_1,x).
\end{align}

Again, the nonlocal and local signals are given by convergent series, while the background is badly divergent. Expanding the summand of the background in large $n$ allows us to identify the divergent pieces in the background as: 
\begin{align}\label{eq_div_DS}
   \text{Div}\Big[\mathcal{J}^{(\nabla\sigma)^2}_{\text{URBG}}\Big] 
  =&\sum_{n=1}^{\infty}\Big[\fr{4n(3+4n)[19+4n(3+2n)]-48n(3+4n)\nn^2+80\nn^4+3(63+88\nn^2)}{256\pi^2n}\mc{T}_1 \n\\
  &~~~~~~-\fr{19+6n+8n^2+12\nn^2}{32\pi^2n}\mc{T}_2 +\fr{1}{8\pi^2n}\mc{T}_3\Big].
\end{align}
Here the summation over $\Delta_{\pm}=3/2\pm\ii\nn$ has been finished, and we have defined three contact shapes:
\begin{align}
\label{eq_T1}
  \mathcal{T}_1=&~\FR{k_s}{k_{1234}},\\
\label{eq_T2}
  \mathcal{T}_2=&~ \FR{k_s}{k_{1234}}+\FR{k_{12}k_{34}k_s}{k_{1234}^3}+\FR{k_s^3}{k_{1234}^3} ,\\
\label{eq_T3}
  \mathcal{T}_3=&~ \FR{k_s}{k_{1234}}+\FR{k_{12}k_{34}k_s}{k_{1234}^3}-\FR{k_s^3}{k_{1234}^3}+\FR{6k_s(k_{12}k_{34}+k_s^2)^2}{k_{1234}^5}. 
\end{align}
Following the standard diagrammatic rules in SK formalism, it is straightforward to verify that these three shapes come respectively from 4-point contact diagrams with the following interactions: 
\begin{align}
\label{eq_Dsigma2counterterms}
  \Delta\mathcal{L}_1= a^4\phi_c^4,&&\Delta\mathcal{L}_2= a^4\phi_c^2\nabla^2(\phi_c^2),&&\Delta\mathcal{L}_3= a^4\nabla^2(\phi_c^2)\nabla^2(\phi_c^2).
\end{align}
These are the expected counterterms to renormalize the 1-loop graph.

With the divergences removed, we can write down the finite and renormalized background as:
\begin{align}
   \mathcal{J}^{(\nabla\sigma)^2}_{\text{BG}}=\mathcal{J}^{(\nabla\sigma)^2}_{\text{URBG}}-\text{Div}\Big[\mathcal{J}^{(\nabla\sigma)^2}_{\text{URBG}}\Big]+C_1\mathcal{T}_1+C_2\mathcal{T}_2+C_3\mathcal{T}_3,
\end{align}
where $C_i$ ($i=1,2,3$) are finite numbers determined by renormalization conditions. 

We can again observe that, when viewing the background as a function of $r_1$ with $x$ fixed, the contact contributions $\mathcal{T}_i$ produce $r_1$ dependences up to $\order{r_1^5}$, showing that all terms with higher powers in $r_1$ are renormalization independent. We can again call them subleading irreducible background. These terms start from $r_1^7$, and it is relatively easy to find an expression for it:
\begin{align}
    \mathcal{J}^{(\nabla\sigma)^2}_{\text{SLIBG}}(r_1,x) = 
  &\sum_{\Delta=\Delta_{\pm}} \sum_{n,\ell=0}^{\infty}\mathop{\mathcal{C}^{(\nabla\sigma)^2}_{\Delta_\pm,n}}\FR{\cos(\pi\Delta)^2}{128\cos(2\pi\Delta)}\FR{(\ell+1)_6}{(\fr{\ell}{2}+\Delta+n-\fr{1}{2})_{4}\,(\fr{\ell}{2}-\Delta-n+1)_4}\n\\
  &\times r_1^7\,(-x)^\ell\,{}_3\text{F}_2\bigg[\bgm1,\fr{\ell}{2}+\fr{7}{2},\fr{\ell}{2}+4\\\fr{\ell}{2}+\Delta+n+\fr{7}{2},\fr{\ell}{2}-\Delta-n+5\edm\bigg|r_1^2\bigg].
\end{align}
This is a very suppressed result in the squeezed limit $r_1\to 0$. On the other hand, similar to the previous case, there are also renormalization independent pieces in the background at lower orders of $r_1$, which has the form $f_1(x)r_1+f_2(x)r_1^3+f_3(x)r_1^5$. Then, we can Laurent expand $f_i(x)$ at $x=-1$ as $f_i(x)=\sum_{m}c_{i,m}(1+x)^m$. Clearly, all contact terms $\mathcal{T}_i$ only contribute terms with $m=-1,-3,-5$. Thus, all other terms represent renormalization independent contributions, which are robust predictions with on-shell data and analyticity.

\subsection{Massive vector bubble}

The last example we consider is the 4-point 1-loop bubble correlator  exchanging a pair of massive spin-1 boson, which is generated from interaction (\ref{eq_int_A}). We can again write the dimensionless correlator $\mathcal{J}^{A^2}(r_1,x)$ as a sum of three pieces:
\begin{equation}
  \mathcal{J}^{A^2}(r_1,x)=\mathcal{J}^{A^2}_{\text{NS}}(r_1,x)+\mathcal{J}^{A^2}_{\text{LS}}(r_1,x)+\mathcal{J}^{A^2}_{\text{BG}}(r_1,x),
\end{equation}
Then:
\begin{align}
\mathcal{J}^{A^2}_{\text{NS}}(r_1,x)=&\sum_{\Delta_{\pm}}\sum_{n=0}^{\infty}\mathop{\mathcal{C}^{A^2}_{\Delta_\pm,n}}\mathbf{I}^{2\Delta_\pm+2n}_{\text{NS}}(r_1,x),\\
\mathcal{J}^{A^2}_{\text{LS}}(r_1,x)=&\sum_{\Delta_{\pm}}\sum_{n=0}^{\infty}\mathop{\mathcal{C}^{A^2}_{\Delta_\pm,n}}\mathbf{I}^{2\Delta_\pm+2n}_{\text{LS}}(r_1,x),\\
\mathcal{J}^{A^2}_{\text{URBG}}(r_1,x)=&\sum_{\Delta_{\pm}}\sum_{n=0}^{\infty}\mathop{\mathcal{C}^{A^2}_{\Delta_\pm,n}}\mathbf{I}^{2\Delta_\pm+2n}_{\text{BG}}(r_1,x),
\end{align}
where $\Delta_\pm=3/2\pm\ii\nn_A$, and the subscript ``URBG" means the unregularized background. 
The spectral coefficient $\mathcal{C}^{A^2}_{\Delta,n}$ is computed from the spectral function according to \eqref{eq_coefficient} and the result can again be expressed in terms of $\mathop{\mathcal{C}^{\sigma^2}_{\Delta,n}}$ in (\ref{eq_Csigma2}), as:
\begin{align}
  \mathcal{C}^{A^2}_{\Delta,n}=\FR{3\Delta^2(\Delta-2)^2+2n(4\Delta-3)(\Delta^2+2n^2-3)+n^2(20\Delta^2-24\Delta-3)+4n^4}{(\Delta-1)^2(\Delta-2)^2}\mathop{\mathcal{C}^{\sigma^2}_{\Delta,n}}.
\end{align}
The nonlocal and local signals have convergent series while the background is again divergent. Following the same procedure as above, we find the divergent part of the background reads: 
\begin{align}
     \text{Div}\Big[\mathcal{J}^{A^2}_{\text{URBG}}\Big]  
 = &\sum_{n=1}^{\infty}\Big[\fr{33+4n(3+4n)\big(-5+4n(3+2n)\big)+168\nn_A^2-48n(3+4n)\nn_A^2+144\nn_A^4}{16\pi^2 n (1+4\nn_A^2)^2}\mc{T}_1\n\\
  &~~~~~~~-\fr{-5+6n+8n^2+12\nn_A^2}{2\pi^2 n (1+4\nn_A^2)^2}\mc{T}_2 +\fr{2}{\pi^2 n (1+4\nn_A^2)^2}\mc{T}_3\Big],
\end{align}
where the sum over $\Delta_{\pm}=3/2\pm\ii\nn_A$ has been finished, and the three shapes $\mathcal{T}_i$ $(i=1,2,3)$ are the same as before, given in (\ref{eq_T1})-(\ref{eq_T3}), showing that these divergences will disappear after we include the three counterterms in (\ref{eq_Dsigma2counterterms}), which is again expected. 
The UV-finite background is again given by 
\begin{equation}
  \mathcal{J}^{A^2}_{\text{BG}} =\mathcal{J}^{A^2}_{\text{URBG}}-\text{Div}\Big[\mathcal{J}^{A^2}_{\text{URBG}}\Big] +C_1'\mathcal{T}_1+C_2'\mathcal{T}_2+C_3'\mathcal{T}_3,
\end{equation}
where again the three coefficients are to be determined by renormalization conditions.

All comments about the renormalization independent part of the background in the previous subsection still apply in this case. So, we can again perform a Laurent expansion to leading terms of the background in $r_1$ to identify the leading irreducible background. On the other hand, there is a very suppressed contribution of subleading irreducible background starting from $r_1^7$ that is relatively easy to extract, which has the following expression:
\begin{align}
 \mathcal{J}^{A^2}_{\text{SLIBG}}(r_1,x)= 
  &\sum_{\Delta=\Delta_{\pm}}\sum_{n,\ell=0}^{\infty}\mathop{\mathcal{C}^{A^2}_{\Delta,n}}\FR{\cos(\pi\Delta)^2}{128\cos(2\pi\Delta)}\FR{(\ell+1)_6}{(\fr{\ell}{2}+\Delta+n-\fr{1}{2})_4\,(\fr{\ell}{2}-\Delta-n+1)_4}\n\\
  &\times r_1^7\,(-x)^\ell\,{}_3\text{F}_2\bigg[\bgm1,\fr{\ell}{2}+\fr{7}{2},\fr{\ell}{2}+4\\\fr{\ell}{2}+\Delta+n+\fr{7}{2},\fr{\ell}{2}-\Delta-n+5\edm\bigg|r_1^2\bigg].
\end{align}

\section{Application to Primordial Bispectrum and Trispectrum}\label{sec_pheno}

Up until now, we have been mostly considering correlators of conformal scalars, which is technically simpler but less relevant to CC phenomenology than that of massless scalars. On the other hand, it is known that a conformal correlator is related in a simple way to a cosmological correlator with external massless scalars or tensors. The relation is streamlined by acting differential operators \cite{Arkani-Hamed:2018kmz,Baumann:2019oyu,Arkani-Hamed:2015bza}. Therefore, we can start from results of the previous section and use these differential operators to obtain cosmological correlators of phenomenological relevance. We spell out the details for inflaton trispectrum and bispectrum in the following. 

\subsection{One-loop trispectrum}

\begin{figure}[t]
  \centering 
  \includegraphics[width=0.3\textwidth]{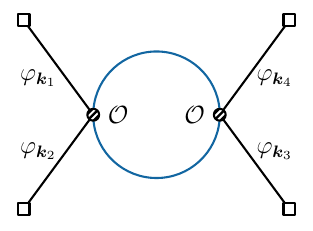} 
  \caption{The 4-point correlator of the massless scalars $\varphi$ mediated by a massive bubble loop exchange. The exchanged fields can be massive scalar $\sigma$ (which can couple to $\varphi_c$ directly or derivatively) or massive vector $A_{\mu}$.}
\label{fd_4pt_bubble_varphi}
\end{figure}

We still consider 1-loop bubble correlators, but now the external modes are massless scalars $\varphi$. In typical inflation models, this massless scalar $\varphi$ is identified as the inflaton fluctuation. In such cases, the approximate shift symmetry of the inflaton dynamics implies that inflaton couplings, if unsuppressed by slow-roll parameters, should involve derivatives. In other models such as curvaton and modulated reheating \cite{Lyth:2001nq,Dvali:2003em,Lu:2019tjj,Kumar:2019ebj}, the external states can couple directly to the massive states. Correlators from all such couplings can be conveniently obtained by acting differential operators on corresponding conformal-scalar correlators. In the following, purely for simplicity, we only consider external states coupled to massive states via a time derivative. Parallel to the three cases in the previous section, we consider three type of interactions. First, the inflaton is coupled to a pair of massive scalar $\si$:  
\begin{equation}\label{eq_int_si_varphi}
    \Delta\mathcal{L}=-\FR{1}4a^2(\varphi')^2\sigma^2.
\end{equation}
Here we use the notation $f'\equiv\partial_\tau f$. Second, the bulk massive scalar $\si$ can also couple to the inflaton through derivatives:
\begin{equation}\label{eq_int_patsi_varphi}
    \Delta\mathcal{L}=-\FR{1}{4}a^2(\varphi')^2(\nabla\sigma)^2.
\end{equation}
Finally, a potentially interesting operator is a bulk massive vector boson coupled to the inflaton:
\begin{equation}\label{eq_int_A_varphi}
    \Delta\mathcal{L}=-\FR{1}{4}a^2(\varphi')^2g^{\mu\nu}A_\mu A_\nu.
\end{equation}
Then the integral representation of the 1-loop correlator can be directly read off from the corresponding diagrammatic rules: 
\begin{align}
    \langle\varphi_{\bm{k}_1}\varphi_{\bm{k}_2}\varphi_{\bm{k}_3}\varphi_{\bm{k}_4}\rangle_{\mathcal{O}}'=&\ \FR{1}{2}\sum_{\mathsf{a},\mathsf{b}=\pm}(-\mathsf{ab})\int_{-\infty}^{0}\FR{\di\tau_1}{(-\tau_1)^2}\FR{\di\tau_2}{(-\tau_2)^2}\big[\partial_{\tau_1}G_{\mathsf{a}}(k_1,\tau_1)\big]\big[\partial_{\tau_1}G_{\mathsf{a}}(k_2,\tau_1)\big]\n \\
    &\times \big[\partial_{\tau_2}G_{\mathsf{b}}(k_3,\tau_2)\big]\big[\partial_{\tau_2}G_{\mathsf{b}}(k_4,\tau_2)\big]\times\mathcal{Q}^{\mathcal{O}}_{\mathsf{ab}}\big(k_s;\tau_1,\tau_2\big),
\end{align}
where $G_{\pm}(k,\tau)$ is the bulk-to-boundary propagator of the inflaton fluctuation $\varphi$: 
\begin{align}
  G_{\aa}(k,\tau)=\FR{1}{2k^3}(1-\ii\aa k\tau)e^{\ii\aa k\tau},
\end{align}
and $\mathcal{Q}^{\mathcal{O}}_{\aa\bb}$ is the 1-loop momentum integral of bulk propagators, whose explicit expressions can be found in (\ref{eq_Q_si}), (\ref{1L4PNaSiNaSiMoIn}) and (\ref{1L4PAAMoIn}).
Again, one can extract a dimensionless seed integral $\mathcal{K}^{\mathcal{O}}$ from the entire correlator, as a function of momentum ratios: 
\begin{align}
    \mathcal{K}^{\mathcal{O}}\Big( \FR{k_s}{k_{12}},\FR{k_{s}}{k_{34}} \Big)\equiv \FR{k_s^5}{2}\!\sum_{\aa,\bb=\pm}(-\aa\bb)\int_{-\infty}^0\!{\di\tau_1}{\di\tau_2}\,\mathcal{Q}^{\mathcal{O}}_{\aa\bb}\big(k_s;\tau_1,\tau_2\big)e^{\ii\aa k_{12}\tau_1+\ii \bb k_{34}\tau_2}.
\end{align}
Here we choose two independent variables as $r_1=k_s/k_{12}$ and $r_2=k_s/k_{34}$. 
Then the seed integral $\mathcal{K}^{\mathcal{O}}$ and the correlator are related by 
\begin{align}
    \langle\varphi_{\bm{k}_1}\varphi_{\bm{k}_2}\varphi_{\bm{k}_3}\varphi_{\bm{k}_4}\rangle_{\mathcal{O}}'=\FR{1}{16k_1k_2k_3k_4k_s^5}\mathcal{K}^{\mathcal{O}}(r_1,r_2).
\end{align} 
Below we focus on the properties of $\mathcal{K}^{\mathcal{O}}$. By comparing $\mathcal{K}^{\mathcal{O}}$ with the previous seed integral $\mathcal{J}^{\mathcal{O}}$ (see (\ref{eq_LoopSeedDef})), we see that the only difference between their definitions is the power of $\tau_{1,2}$ in their integrands. 
Furthermore, $\mathcal{K}^{\mathcal{O}}$ can be obtained by performing appropriate differential operators on $\mathcal{J}^{\mathcal{O}}$: 
\begin{align}
  \mathcal{K}^{\mathcal{O}}(r_1,r_2)=(r_1^4\partial^2_{r_1}+2r_1^3\partial_{r_1})(r_2^4\partial^2_{r_2}+2r_2^3\partial_{r_2})\mathcal{J}^{\mathcal{O}}\Big(r_1,\FR{r_1}{r_2}\Big).
\end{align}
With $\mathcal{J}^{\mathcal{O}}$ obtained in the last section, the full explicit results of $\mc{K}^{\mc{O}}$ are obtained directly. 

\begin{figure}[htbp]
  \centering 
  \includegraphics[width=1\textwidth]{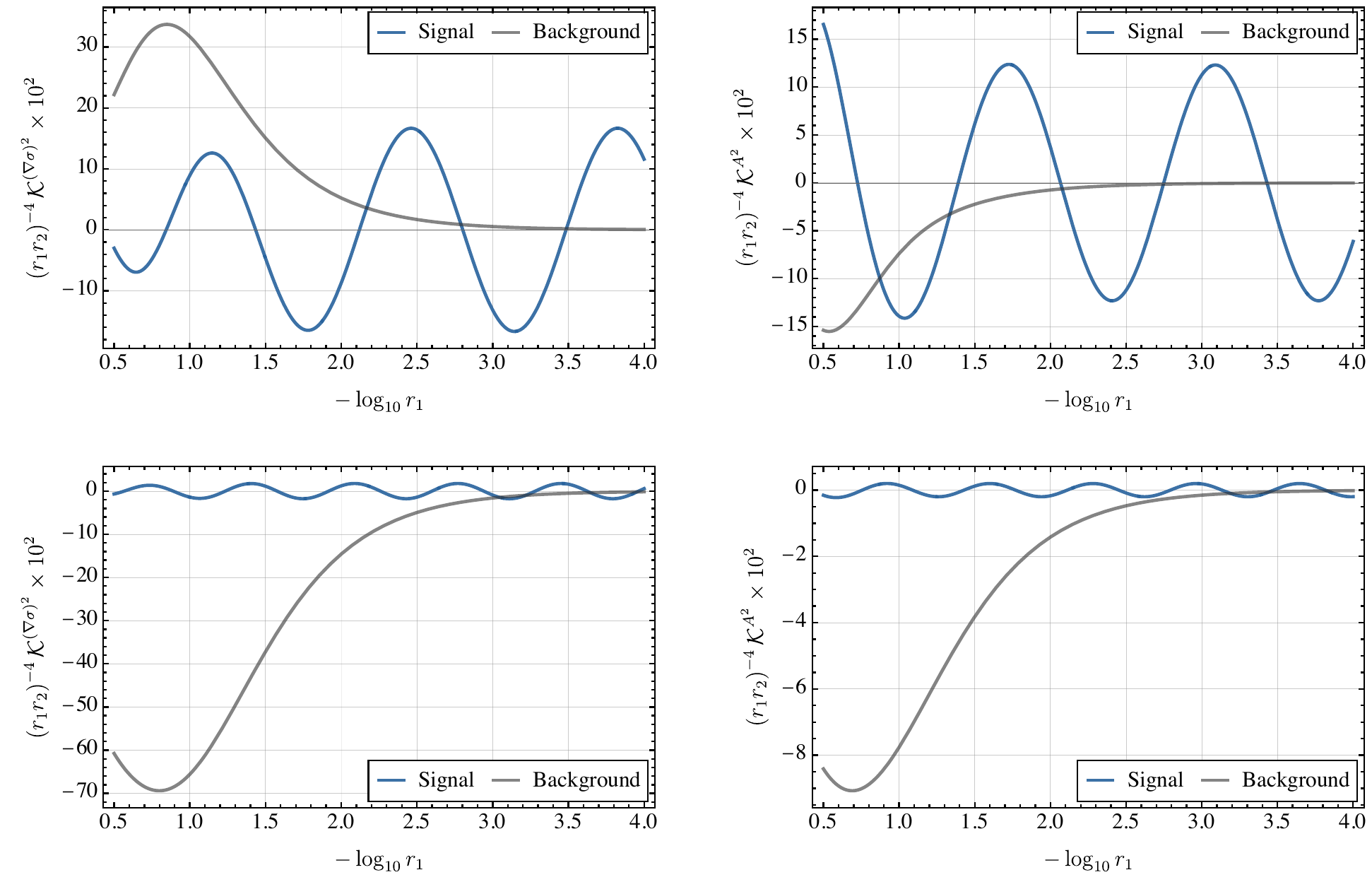} 
  \caption{The 1-loop trispectra $\mathcal{K}^{(\nabla\sigma)^2}$ (left panels) and $\mathcal{K}^{A^2}$ (right panels) as functions of $r_1=k_s/k_{12}$, with $r_2=k_s/k_{34}=0.9$. The upper two panels show the result with $\wt\nu=\wt\nu_A=1$ and the lower two show $\wt\nu=\wt\nu_A=2$.
    The signal curves in dark blue collects both nonlocal and local contributions, $\mathcal{K}^{\mathcal{O}}_{\text{S}}\equiv\mathcal{K}^{\mathcal{O}}_{\text{NS}}+\mathcal{K}^{\mathcal{O}}_{\text{LS}}$; The background curves, in gray, show the renormalized results with modified minimal subtraction detailed in the main text.}
\label{fig_trispectra}
\end{figure}

Instead of showing the explicit results of $\mc{K}^{\mc{O}}$, we plot their shapes in Fig.\;\ref{fig_trispectra} for $\mc{O}=(\nabla\si)^2$ and $A^2$. (Similar plots for $\mc{O}=\si^2$ can be found in \cite{Xianyu:2022jwk}). For clarity, we fix $r_2=0.9$ and plot the trispectra as functions of $r_1$. Also, we plot the signal (both local and nonlocal) and the background separately. Here, we have shown the regularized background with a divergent part subtracted. For the finite part, we adopted a ``modified minimal subtraction.'' More precisely, by expressing the background as a sum of quasinormal modes $\mc J=\sum_n \mc C_n \mb I^n$, we expand the summand in large $n$. Then, we remove all divergent parts such as $c_{-1}\sum_{n} / n$, $c_0\sum_n 1$, $c_1\sum_n n$, $\cdots$. In addition, we also remove a convergent term from the sum $ c_{-2}\sum_{n=1}^\infty /n^2=c_{-2}\pi^2/6$. It happens that all coefficients $c_\ell$ with $\ell\geq -2$ have kinematic dependences identical to the expected counterterms. The resulting background shown in the Figure thus contains a subtraction-dependent piece which dominates the background in the squeezed limit.

\paragraph{Squeezed and large-mass limits} 
Given the long expressions for all full shapes computed above, it is useful to show their approximated form in the squeezed limit, where the signal is most prominent, and also the large mass limit, where the parametric dependence of the mass has a more transparent form. 

For trispectra, we consider a hierarchical squeezed limit $r_1\ll r_2\ll1$ as in \cite{Xianyu:2022jwk}. In this limit, the signals, both nonlocal and local, are dominated by quasinormal modes of smallest scaling dimensions. The background, on the other hand, is dominated by a counterterm contribution which is not to be determined from our calculation. Here, we show both the leading-order counterterm contribution with an undetermined coefficient and the leading-order irreducible background that is independent of renormalization conditions. 

The hierarchical squeezed limit of the trispectrum from the derivatively coupled scalar bubble is:
\begin{align}
  \lim_{r_1\ll r_2\ll1}\mc K^{(\nabla\si)^2}_\rm{NS}(r_1,r_2)=&~\FR{(1-\ii\nn)(3-2\ii\nn)^2\mathop{\rm{sech}}(2\pi\nn)}{4^4\pi^2}\FR{\Gamma^4(\fr{5}{2}-\ii\nn)}{\Gamma(4-4\ii\nn\big)} (4r_1r_2)^{4-2\ii\nn}+\rm{c.c.},
  \\\lim_{r_1\ll r_2\ll1}\mc K^{(\nabla\si)^2}_\rm{LS}(r_1,r_2)=&~\FR{(3-2\ii\nn)^4\mathop{\rm{sech}}(2\pi\nn)}{32\sqrt{\pi}}\FR{(1-\ii\nn)_{1/2}(-\ii\nn)_{5/2}}{(2\ii\nn)_{-1/2}} r_1^5\Big(\FR{r_1}{r_2}\Big)^{-1-2\ii\nn}+\rm{c.c.}.
\end{align}
The leading background has a fixed dependence on the momentum ratio in the squeezed limit but the coefficient $C^{(\nabla\si)^2}_{\rm{LBG}}$ depends on renormalization condition, which we choose not to fix here:
\begin{align}
  \lim_{r_1\ll r_2\ll1}\mc K^{(\nabla\si)^2}_{\rm{LBG}}=C^{(\nabla\si)^2}_{\rm{LBG}} r_1^5.
\end{align}
On the other hand, the leading renormalization-independent background comes from the SLIBG defined before:
\begin{align}
  \lim_{r_1\ll r_2\ll1}\mc K^{(\nabla\si)^2}_\rm{SLIBG}(r_1,r_2)=\bigg[\sum_{n=0}^{\infty}\FR{14175\big(1-\mathop{\rm{sech(2\pi\nn)}}\big)}{(2+n-\ii\nn)_4(\fr{1}{2}-n+\ii\nn)_4}\mc C^{(\nabla\si)^2}_{3/2-\ii\nn,n}\bigg] r_1^{11}+\rm{c.c.}.
\end{align}  
Here the coefficient itself is a $\wt\nu$-dependent pure number computed from a convergent summation.

The limiting behavior for the massive vector bubble is:
\begin{align}
  \lim_{r_1\ll r_2\ll1}\mc K^{A^2}_\rm{NS}(r_1,r_2)=&~\FR{3(1-\ii\nn_A)\mathop{\rm{sech}}(2\pi\nn_A)}{16\pi^2(1-2\ii\nn_A)^2}\FR{\Gamma[\fr{5}{2}-\ii\nn_A]^4}{\Gamma[4-4\ii\nn_A]} (4r_1 r_2)^{4-2\ii\nn_A}+\rm{c.c.},
  \\\lim_{r_1\ll r_2\ll1}\mc K^{A^2}_\rm{LS}(r_1,r_2)=&~\FR{3(3-2\ii\nn_A)^2\mathop{\rm{sech}}(2\pi\nn_A)}{2\sqrt{\pi}(1-2\ii\nn_A)^2}\FR{(1-\ii\nn_A)_{1/2}(-\ii\nn_A)_{5/2}}{(2\ii\nn_A)_{-1/2}} r_1^5\Big(\FR{r_1}{r_2}\Big)^{-1-2\ii\nn_A}+\rm{c.c.},
  \\\lim_{r_1\ll r_2\ll1}\mc K^{A^2}_\rm{LBG}(r_1,r_2)\propto&~ r_1^5,
  \\\lim_{r_1\ll r_2\ll1}\mc K^{A^2}_\rm{SLIBG}(r_1,r_2)=&~\sum_{n=0}^{\infty}\FR{14175\big(1-\mathop{\rm{sech(2\pi\nn_A)}}\big)}{(2+n-\ii\nn_A)_4(\fr{1}{2}-n+\ii\nn_A)_4}\mc C^{A^2}_{3/2-\ii\nn_A,n} r_1^{11}+\rm{c.c.}.
\end{align} 

Starting from the above limiting expressions, it is useful to further take the large mass limit $\wt\nu\gg 1$ and $\wt\nu_A\gg 1$ and the above trispectra further simplify to the following. For the derivatively coupled scalar bubble, we have:
\begin{align}
  &\lim_{\nn\gg1}\lim_{r_1\ll r_2\ll1}\mc K^{(\nabla\si)^2}_\rm{NS}(r_1,r_2)=\FR{2^{11/2}}{\sqrt{\pi}}e^{{\ii\pi}/{4}}\nn^{{15}/{2}}e^{-2\pi\nn}\times\Big(\FR{r_1 r_2}{4}\Big)^{4-2\ii\nn}+\rm{c.c.},
  \\&\lim_{\nn\gg1}\lim_{r_1\ll r_2\ll1}\mc K^{(\nabla\si)^2}_\rm{LS}(r_1,r_2)=\sqrt{\FR{2}{\pi}}e^{{3\ii\pi}/{4}}\nn^{{15}/{2}}e^{-2\pi\nn}\times r_1^5\Big(\FR{r_1}{r_2}\Big)^{-1-2\ii\nn}+\rm{c.c.}.
\end{align}
The meaning of large-mass limit for the leading background is unclear since it depends on the renormalization condition, so we do not consider it. On the other hand, the subleading renormalization-independent part has the following large-mass limit:
\begin{align}
 \lim_{\nn\gg1}\lim_{r_1\ll r_2\ll1}\mc K^{(\nabla\si)^2}_\rm{SLIBG}(r_1,r_2)=\FR{6210}{\pi^2}\FR{r_1^{11}}{\nn^2}.
\end{align}
For the vector bubbles, we have:
\begin{align}
  &\lim_{\nn_A\gg1}\lim_{r_1\ll r_2\ll1}\mc K^{A^2}_\rm{NS}(r_1,r_2)=\FR{3\times2^{11/2}}{\sqrt{\pi}}e^{{\ii\pi}/{4}}\nn_A^{{7}/{2}}e^{-2\pi\nn_A}\times \Big(\FR{r_1r_2}{4}\Big)^{4-2\ii\nn_A}+\rm{c.c.},
  \\&\lim_{\nn_A\gg1}\lim_{r_1\ll r_2\ll1}\mc K^{A^2}_\rm{LS}(r_1,r_2)=\FR{6}{\sqrt{2\pi}}e^{{3\ii\pi}/{4}}\nn_A^{{7}/{2}}e^{-2\pi\nn_A}\times r_1^5\Big(\FR{r_1}{r_2}\Big)^{-1-2\ii\nn_A}+\rm{c.c.},
  \\&\lim_{\nn_A\gg1}\lim_{r_1\ll r_2\ll1}\mc K^{A^2}_\rm{SLIBG}(r_1,r_2)=\FR{7290}{\pi^2}\FR{r_1^{11}}{\nn_A^6}.
\end{align}

\subsection{One-loop bispectrum}

The bispectra with massive bubble exchanges can  be obtained similarly. 
An example has been given in Sec.\;\ref{sec_dispersion_method_3pt}, where we considered the bispectrum mediated by a pair of directly coupled scalars (See interaction (\ref{eq_int_3pt_si}) and Fig.\;\ref{fd_3pt_bubble}). A general lesson there is that it is advantageous to use the variable $u\equiv2k_3/k_{123}$ to express the bispectrum. For one thing, the UV divergence appears as a local term $u^3$, which is exactly the leading order of the background in the small $u$ expansion. This makes it easier to identify and to remove the UV divergence of a 3-point function than the 4-point case.

In this section we study the derivatively coupled scalar bubble and the massive vector bubble, and give their bispectra. The procedure is the same as the 4-point case but the results are simpler. Here, we only show the final results.

\paragraph{Full results}  
The full result (including both signal and background) for the derivatively coupled scalar bubble reads: 
\begin{align}
\label{eq_K3ptNablaSigma2}
  &\mathcal{K}^{(\nabla\sigma)^2}_{\text{3-pt}}(u) \n\\
  &=\sum_{\Delta=\Delta_\pm}\sum_{n=0}^{\infty}\mathcal{C}^{(\nabla\sigma)^2}_{\Delta,n}\bigg\{\mathbf{I}_{\text{3-pt,\,S}}^{2\Delta+2n}(u)+\FR{\pi}{4}\FR{\cos(2\pi\Delta)+1}{\sin(4\pi\Delta)}u^6\,{}_3\mathcal{F}_2\bigg[\bgm1,4,6\\2\Delta+2n+3,-2\Delta-2n+6\edm\bigg|u\bigg]\bigg\}\n\\
  &~~~+C_3 u^3+C_4 u^4+C_5 u^5.
\end{align}
Here $\mathbf{I}^{\Delta}_{\text{3-pt,\,S}}$ is the signal of the 3-point function with single quasinormal-mode exchange, given in (\ref{eq_tree_3ptS_QNMode}), and the sum of these functions contributes to the 1-loop signal, starting from $\order{u^{4\pm2\ii\nn}}$ in the small $u$ limit. The renormalization-independent background includes terms at $\mathcal{O}(u^6)$ and at higher orders. The more leading orders in the background, including $\order{u^3}$, $\order{u^4}$, and $\order{u^5}$, come from local counterterms, with coefficients $C_3$, $C_4$, and $C_5$ to be determined by renormalization conditions.
 
The full result for the massive vector bubble is: 
\begin{align}
\label{eq_K3ptA2}
  &\mathcal{K}^{A^2}_{\text{3-pt}}(u)\n\\
  &=\sum_{\Delta_\pm}\sum_{n=0}^{\infty}\mathcal{C}^{A^2}_{\Delta_{\pm},n}\bigg\{\mathbf{I}_{\text{3-pt,\,S}}^{2\Delta_\pm+2n}(u)+\FR{\pi}{4}\FR{\cos(2\pi\Delta_\pm)+1}{\sin(4\pi\Delta_\pm)}u^6\,{}_3\mathcal{F}_2\bigg[\bgm1,4,6\\2\Delta_\pm+2n+3,-2\Delta_\pm-2n+6\edm\bigg|u\bigg]\bigg\}\n\\
  &~~~+C_{3}' u^3+C_{4}' u^4+C_{5}' u^5.
\end{align}
Again, the $\order{u^3}$ to $\order{u^5}$ of the background receive contributions from local counterterms with coefficients determined by renormalization conditions. 

\begin{figure}[htbp]
  \centering 
  \includegraphics[width=1\textwidth]{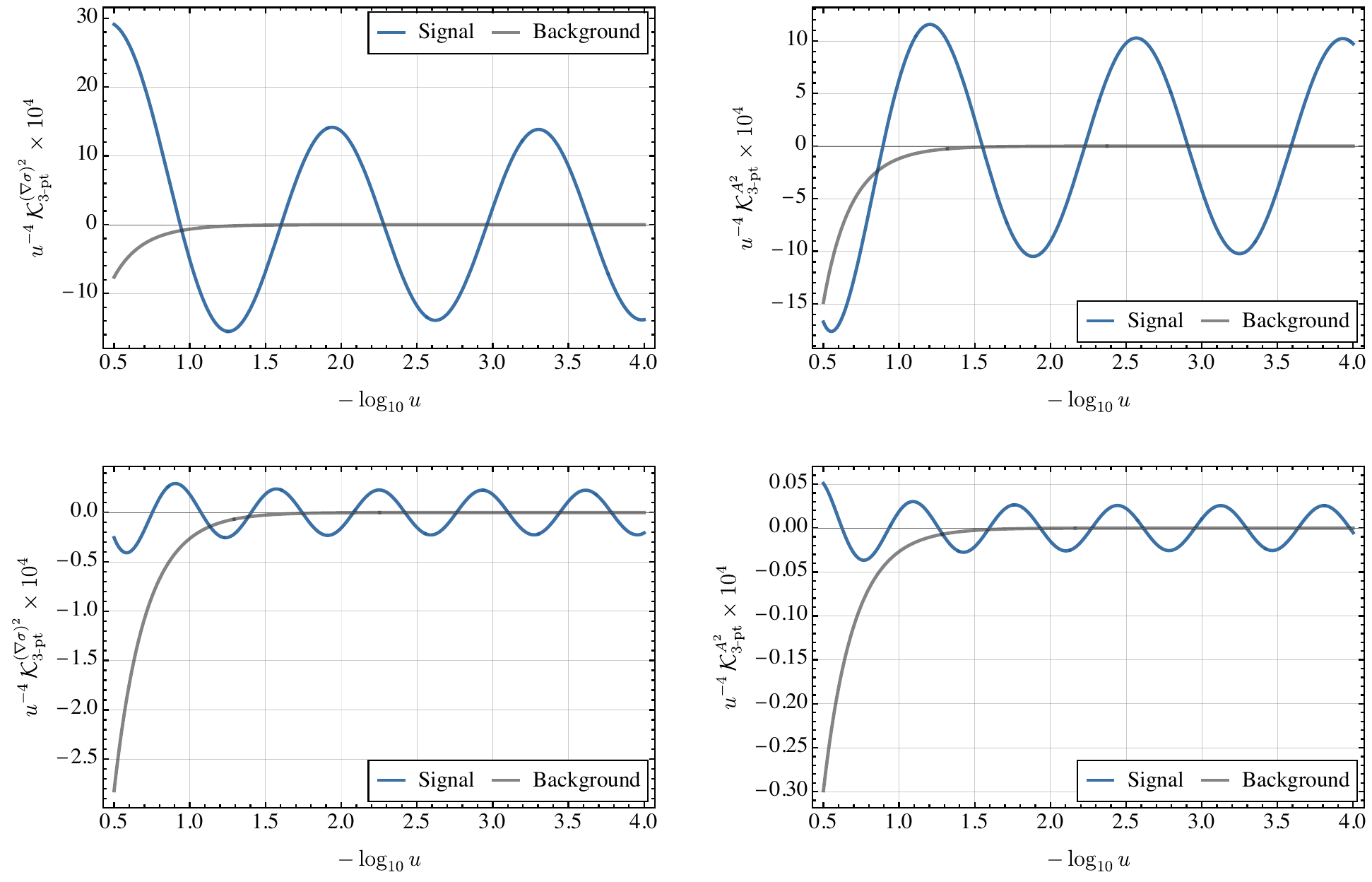} 
  \caption{The 1-loop bispectra $\mathcal{K}^{(\nabla\sigma)^2}_{\text{3-pt}}$ (left two panels) and $\mathcal{K}^{A^2}_{\text{3-pt}}$ (right two panels) as functions of $u\equiv 2k_3/k_{123}$. The two upper panels show $\wt\nu=\wt\nu_A=1$ and the lower two show $\wt\nu=\wt\nu_A=2$. }
  \label{fig_bispectra}
\end{figure}

The 1-loop bubble bispectra from $\mathcal{K}^{(\nabla\sigma)^2}_{\text{3-pt}}$ and $\mathcal{K}^{A^2}_{\text{3-pt}}$ are shown in Fig.\;\ref{fig_bispectra}. Here, we choose to show the renormalization-independent part of the background. For this purpose, we set all $C_{3,4,5}$ in (\ref{eq_K3ptNablaSigma2}) and $C_{3,4,5}'$ in (\ref{eq_K3ptA2}) to 0. So, Fig.\;\ref{fig_bispectra} should not be interpreted as that the signal is dominating over the background; Rather, the signal is dominating only over the irreducible background. When renormalization-dependent terms are included, the background is still dominant in the squeezed limit.

\paragraph{Squeezed and large-mass limits} Similar to trispectra, it is again useful to consider the squeezed limit and the large-mass limit of the correlators. In the squeezed limit $u\ll1$, the signal part of derivative scalar bubble correlator reads:  
\begin{align}
  \lim_{u\ll1}\mc K^{(\nabla\si)^2}_{\text{3-pt,\,S}}(u)=-\ii\times2^{-3}(3-2\ii\nn)^4\mathop{\rm{csch}}(2\pi\nn)\FR{(1-\ii\nn)_{{3}/{2}}}{(\fr{3}{2}-\ii\nn)_{-{1}/{2}}}\times \Big(\FR{u}{4}\Big)^{4-2\ii\nn}+\rm{c.c.}.
\end{align}
The leading background is directly read off from \eqref{eq_K3ptNablaSigma2} as
\begin{align}
  \lim_{u\ll1}\mc K^{(\nabla\si)^2}_{\text{3-pt,\,LBG}}(u)=\mc C_3u^3,
\end{align}
where $\mc C_3$ depends on the renormalization condition. On the other hand, the leading renormalization-independent background part is:
\begin{align}
  \lim_{u\ll1}\mc K^{(\nabla\si)^2}_{\text{3-pt,\,SLIBG}}(u)=\sum_{n=0}^{\infty}\FR{90\big(\mathop{\rm{sech}}(2\pi\nn)-1\big)}{(-2+2n-\ii\nn)_8}\mc C^{(\nabla\si)^2}_{3/2-\ii\nn,n}\times u^6+\rm{c.c.},
\end{align}
and the vector-bubble bispectrum reads: 
\begin{align}
  \lim_{u\ll1}\mc K^{A^2}_{\text{3-pt,\,S}}(u)=&~-\FR{6\ii(3-2\ii\nn_A)^2\mathop{\rm{csch}}(2\pi\nn_A)}{(1-2\ii\nn_A)^2}\FR{(1-\ii\nn_A)_{{3}/{2}}}{(\fr{3}{2}-\ii\nn_A)_{-{1}/{2}}}\times \Big(\FR u 4\Big)^{4-2\ii\nn_A}+\rm{c.c.},
  \\\lim_{u\ll1}\mc K^{A^2}_{\text{3-pt,\,LBG}}(u)=&~ C_3' u^3,
  \\\lim_{u\ll1}\mc K^{A^2}_{\text{3-pt,\,SLIBG}}(u)=&~\sum_{n=0}^{\infty}\FR{90\big(\mathop{\rm{sech}}(2\pi\nn_A)-1\big)}{(-2+2n-\ii\nn_A)_8}\mc C^{A^2}_{3/2-\ii\nn_A,n}\times u^6+\rm{c.c.}.
\end{align}

We can also consider the limit with both $u\ll1$ and $\nn\gg1$. For the bispectrum with a derivatively coupled scalar bubble, we have,
\begin{align}
  &\lim_{\nn\gg1}\lim_{u\ll1}\mc K^{(\nabla\si)^2}_{\text{3-pt,\,S}}(u)=4e^{{\ii\pi}/{2}}\nn^6e^{-2\pi\nn}\times \Big(\FR u 4\Big)^{4-2\ii\nn}+\rm{c.c.},
  \\&\lim_{\nn\gg1}\lim_{u\ll1}\mc K^{(\nabla\si)^2}_{\text{3-pt,\,SLIBG}}(u)=-\FR{69}{448\pi^2}\FR{u^6}{\nn^2}.
\end{align}
For the vector bubble, we have
\begin{align}
  &\lim_{\nn_A\gg1}\lim_{u\ll1}\mc K^{A^2}_{\text{3-pt,\,S}}(u)=3\times12e^{{\ii\pi}/{2}}\nn_A^2e^{-2\pi\nn_A}\times \Big(\FR u 4\Big)^{4-2\ii\nn_A}+\rm{c.c.},
  \\&\lim_{\nn_A\gg1}\lim_{u\ll1}\mc K^{A^2}_{\text{3-pt,\,SLIBG}}(u)=-\FR{81}{448\pi^2}\FR{u^6}{\nn_A^6}.
\end{align}
Same as the 4-point case, since the leading background part depends on the renormalization condition, we do not consider its large mass limit.

\section{Conclusions and Outlooks}\label{sec_conclusion}

Cosmological collider physics has experienced rapid development in recent years, with numerous new questions, methodologies, and results emerging from both theoretical and observational perspectives. After a decade of development, the field has evolved from a purely theoretical curiosity to an active intersection of observational cosmology, particle phenomenology, and amplitude communities. With the current and future observational data, many new opportunities are now open for us to make real progress. 

Inflation correlators with massive exchanges are certainly central to the study of CC physics. To confront real data, high-quality template banks motivated by sound CC models are desperately needed, making the fast and precise numerical implementation of massive correlators a problem more pressing than ever. In this regard, an analytical approach to inflation correlators is certainly very helpful. 

The development on the analytical front showed that massive correlators are indeed very complicated objects. For one thing, they involve many unfamiliar special functions. These results give us a feeling that it is simply not enough to just find analytical expressions for a process; We need to understand them. Can we represent them in various physically transparent or computationally efficient forms? Can we fully decode their analytical properties? Can we use their (non)analyticity as a probe of bulk dS physics or as an efficient bootstrap tool? We think that these are the current big questions on the analytical frontier that should be addressed.

In this work, we put forward an idea of on-shell bootstrap of massive correlators based on their peculiar analytic properties at tree and loop levels. Conceptually, there is nothing terribly new in our approach; All we have done is merely to show that an appropriate combination of known techniques, including dispersion relations and spectral decomposition, coupled with certain insights into the analyticity of massive correlators, can lead to great simplifications for bootstrapping loop processes. 

Since we use the nonlocal signal as the only input, which ultimately comes from the on-shell massive particles propagating in the bulk dS, our approach echoes the well-developed on-shell program of scattering amplitudes in flat space. As such, it shares the advantages of typical on-shell techniques, including a neat separation of on-shell propagating effects from local EFT counterparts demanded by analyticity, and, within the EFT background, a neat separation of UV-sensitive local counterterms from renormalization-independent pieces. Also, our results make manifest the intuitive picture that a process involving an intermediate massive loop can be viewed as a discrete superposition of decaying modes with definite scaling dimensions. When applied to previously studied processes, our method leads to much simplified results. Meanwhile, we also showcase its use for more complicated (but phenomenologically relevant) processes, including derivatively coupled and spinning massive loops. In these cases, our analytical results are new.

Our results here only represent a first use of spectral dispersion. There are certainly many open problems to be studied along this direction, some of which are trivial, but others are not. We list several of them to conclude this paper. 

First, there are a few technically straightforward extensions that may nevertheless be of interest from the viewpoints of CC model buildings. They include various types of massive loops, such as bubbles involving states of different masses, fermions, higher-spin particles, other higher-derivative couplings, and nonzero total angular momentum. Essentially, spectral decomposition is applicable as long as the dispersions of the bubble lines are dS covariant. Importantly, this covers the situation where the interaction itself is noncovariant, as we discussed in Sec.\;\ref{sec_spectral} around (\ref{eq_LagPhi2SigmaPrime2}). The same is true for multiloop banana graphs with dS-covariant loop lines and noncovariant vertices. We can also couple these bubble/banana functions to external states with more exotic properties, such as an inflaton with nonunit sound speed.

Second, there are natural follow-ups along this direction, which, however, may be relatively nontrivial to carry out. One such problem is to seek a similar representation for massive bubbles that break (badly) some of the dS isometries. A particularly interesting scenario is a massive loop with fields of nonzero spin and nonzero helical chemical potential. This is a highly interesting situation for CC physics, but notoriously hard to compute. The spectral decomposition in its original form is unlikely in this case, yet we expect that the meromorphicity of the quasinormal-mode spectrum and analytic structure on the complex momentum space remain the same. So, there may be a hope to bootstrap such processes with techniques not heavily relying on dS symmetry. Also, there are other topologies at 1-loop order, such as triangle and box graphs, that are equally interesting for CC applications. Conceptually, the on-shell strategy should still work for these processes. The difficulty, however, is the increasing number of independent kinematic variables, which complicate  the analytical structure in the multidimensional kinematic space. We leave all these interesting and challenging problems for future studies.

\paragraph{Acknowledgments} We thank Kezhao Pan, Dong-Gang Wang, Masahide Yamaguchi, and Yuhang Zhu for useful discussions. This work is supported by NSFC under Grants No.\ 12275146 and No.\ 12247103, the National Key R\&D Program of China (2021YFC2203100), and the Dushi Program of Tsinghua University.

\begin{appendix}

\section{Useful Functions}\label{appd_function}

For reader's convenience, in this appendix we list some special functions used in the main text.

\paragraph{Gamma functions} In this work we use the following notations to represent the products and ratios of Euler $\Gamma$ functions:
\begin{align}
  \Gamma\big[a_1,a_2,\cdots,a_n\big]\equiv\Gamma(a_1)\Gamma(a_2)\cdots\Gamma(a_n),
\end{align}
\begin{align}
  \Gamma\bigg[\bgm a_1,a_2\cdots,a_n\\b_1,b_2\cdots,b_n\edm\bigg]\equiv\FR{\Gamma(a_1)\Gamma(a_2)\cdots\Gamma(a_n)}{\Gamma(b_1)\Gamma(b_2)\cdots\Gamma(b_n)}.
\end{align}
We also heavily use the Pochhammer symbol $(a)_n$, which is defined as 
\begin{align}
  (a)_n\equiv\FR{\Gamma(a+n)}{\Gamma(a)}.
\end{align}

\paragraph{Hypergeometric functions} The (generalized) hypergeometric functions frequently appear in our results of inflation correlators. The standard form of the hypergeometric function ${}_p\text{F}_q[\cdots]$ is defined by the following series within the convergence region, and by analytical continuation elsewhere:
\begin{align}
  {}_p\text{F}_q\bigg[\bgm a_1,a_2,\cdots,a_p\\b_1,b_2,\cdots,b_q\edm\bigg|z\bigg]\equiv\sum_{n=0}^{\infty}\FR{(a_1)_n(a_2)_n\cdots(a_p)_n}{(b_1)_n(b_2)_n\cdots(b_q)_n}\FR{z^n}{n!}.
\end{align}
In this work we also define the ``dressed'' version of hypergeometric ${}_p\mathcal{F}_q[\cdots]$, which is defined as: 
\begin{align}
  {}_p\mathcal{F}_q\bigg[\bgm a_1,a_2,\cdots,a_p\\b_1,b_2,\cdots,b_q\edm\bigg|z\bigg]=\Gamma\bigg[\bgm a_1,a_2,\cdots,a_p\\b_1,b_2\cdots,b_q\edm\bigg]{}_p\text{F}_q\bigg[\bgm a_1,a_2,\cdots,a_p\\b_1,b_2,\cdots,b_q\edm\bigg|z\bigg].
\end{align}

\section{Spectral Decomposition and Spectral Functions}\label{appd_spectral}

In this appendix, we compute the spectral functions we used in the main text, including $\rho^{\sigma^2}(\wt{\nu}')$, $\rho^{(\nabla\sigma)^2}(\wt{\nu}')$, and $\rho^{A^2}(\wt{\nu}')$. Our calculation closely follows \cite{Loparco:2023rug}, where the explicit expression for $\rho^{\sigma^2}(\wt{\nu}')$ was derived. The corresponding expressions for $\rho^{(\nabla\sigma)^2}(\wt{\nu}')$ and $\rho^{A^2}(\wt{\nu}')$ can be found in \cite{Zhang:2025nzd}, where similar calculations were performed.

\subsection{Embedding Formalism}

The computation is conveniently performed in embedding space\cite{Loparco:2023rug,Costa:2014kfa}, where a $(d+1)$-dimensional de Sitter spacetime can be considered as a hypersurface in a $(d+2)$-dimensional Minkowski spacetime:
\begin{equation}
    -Y_0^2+Y_1^2+\cdots+Y_{d+1}^2=1,
\end{equation}
and in particular, the Poincar\'e patch of dS$_{d+1}$ is parametrized as:
\begin{equation}
    Y^0=-\frac{1-\tau^2+\mathbf{y}^2}{2\tau},\quad Y^i=-\frac{y^i}\tau,\quad Y^{d+1}=-\frac{1+\tau^2-\mathbf{y}^2}{2\tau}.
\end{equation}

There is a natural one-to-one correspondence between symmetric traceless tensors on $\mathrm{dS}_{d+1}$, $T_{\mu_1\cdots\mu_J}(y)$, and symmetric traceless embedding-space tensors restricted to $Y^2=1$ and satisfying the tangentiality condition, $T_{A_1\cdots A_J}(Y)$. In particular, the correspondence is given by the following projection:
\begin{equation}
    T_{\mu_1\cdots\mu_J}(y)
    =
    \frac{\partial Y^{A_1}}{\partial y^{\mu_1}}
    \cdots
    \frac{\partial Y^{A_J}}{\partial y^{\mu_J}}
    T_{A_1\cdots A_J}(Y),
    \qquad
    Y^{A_1}T_{A_1\cdots A_J}(Y)=0.
\end{equation}
Furthermore, it would be convenient to represent the tensor in the embedding space in an index-free way by contracting its indices with a null vector $W^A$:
\begin{equation}
    T(Y,W)=W^{A_1}\cdots W^{A_J}T_{A_1\cdots A_J}(Y),\qquad W^2=0.
\end{equation}
As the tensor $T_{A_1\cdots A_J}(Y)$ is tangential, we can further require that the null vector $W^A$ is also tangential, i.e., $Y\cdot W=0$. To restore the tensor from its index-free representation, one just need to act the following differential operator on $T(Y, W)$:
\begin{align}\label{KVector}
    K_A=&~\bigg(\frac{d-1}2\bigg)\bigg[\frac{\partial}{\partial W^A}-Y_A\bigg(Y\cdot\frac{\partial}{\partial W}\bigg)\bigg]+\bigg(W\cdot\frac{\partial}{\partial W}\bigg)\frac{\partial}{\partial W^A}-Y_A\bigg(Y\cdot\frac{\partial}{\partial W}\bigg)\bigg(W\cdot\frac{\partial}{\partial W}\bigg)\n \\
    &-\frac12W_A\bigg[\frac{\partial^2}{\partial W\cdot\partial W}-\bigg(Y\cdot\frac{\partial}{\partial W}\bigg)\bigg(Y\cdot\frac{\partial}{\partial W}\bigg)\bigg],
\end{align}
which is defined to be interior to the submanifold $Y^2-1=W\cdot Y=W^2=0$. Similarly, the embedding differential operator becomes:
\begin{equation}\label{eq_nabla}
    \nabla_A=\frac{\partial}{\partial Y^A}-Y_A\bigg(Y\cdot\frac{\partial}{\partial Y}\bigg)-W_A\bigg(Y\cdot\frac{\partial}{\partial W}\bigg).
\end{equation}

\subsection{Inversion Formulae}

Without loss of generality, we restrict our discussions to the scalar case. The Wightman function of a scalar operator $\mathcal O$ in dS space admits the following K\"all\'en-Lehmann representation in the embedding formalism for $d\geq 3$ \cite{Loparco:2023rug}:
\begin{align}\label{eq_OKL}
    D^{\mathcal{O}}_{-+}(Y_1,Y_2)=&\int_{-\infty}^\infty\mathrm{d}\wt{\nu}'\rho^{\mathcal{O}}(\wt{\nu}')D_{-+}^{\wt{\nu}'}(Y_1,Y_2),
\end{align}
where $D_{-+}^{\wt{\nu}'}$ is the Wightman function of a free scalar field with mass parameter $\wt\nu'$ (see \eqref{eq_propagator_si}), and $\rho^{\mathcal{O}}(\wt{\nu}')$ is the spectral function of $\mathcal O$.
Since $\rho^{\mathcal O}(\wt{\nu}')$ is real when $\wt{\nu}'$ is real, all the four bulk-to-bulk propagators of $\mathcal O$ share the similar representation with the same spectral function $\rho^{\mathcal O}(\wt{\nu}')$.

To compute the spectral function, it would be useful to Wick rotate to Euclidean anti-de Sitter (EAdS) space, where harmonic analysis play an important role. Given the Bunch-Davis initial condition, the Wick rotation of the coordinates is given by:\footnote{Although not manifest here, we should also Wick rotate $H^2\to -L^2$. After that, we set $L=1$ for simplicity. Therefore, the constraint $Y^2=1$ for dS becomes $X^2=-1$ for EAdS.}
\begin{equation}
    Y_1(\tau_1,\mathbf{y}_1)\xrightarrow{\tau_1\rightarrow\tau_1e^{\ii\pi/2}}X_1(\tau_1,\mathbf{y}_1),\quad Y_2(\tau_2,\mathbf{y}_2)\xrightarrow{\tau_2\rightarrow\tau_2e^{-\ii\pi/2}}X_2(\tau_2,\mathbf{y}_2).
\end{equation}
The Wightman function of a free scalar field with mass parameter $\wt\nu$ in dS then becomes
\begin{equation}\label{D2Omega}
    D_{-+}^{\wt{\nu}}(Y_1,Y_2)\rightarrow\Gamma[\ii\wt{\nu},-\ii\wt{\nu}]\Omega^{\wt{\nu}}(X_1,X_2)
\end{equation}
in EAdS, where the harmonic function is well known to be (see e.g. \cite{Loparco:2023rug,Costa:2014kfa}):
\begin{equation}\label{eq_Omega}
    \Omega^{\lambda}(X_1,X_2)= \frac{1}{2^{d+1}\pi^{(d+1)/2}\Gamma[\ii\lambda,-\ii\lambda]}\,{}_2\mathcal F_1\left[\begin{matrix}\frac d2+\ii\lambda,\frac d2-\ii\lambda \\ \frac{d+1}2 \end{matrix}\middle| \frac{1+X_1\cdot X_2}{2}\right].
\end{equation}

An important feature of the harmonic function \eqref{eq_Omega} is that they are orthogonal in the following sense:
\begin{equation}\label{eq_ortho}
    \int_X\Omega^{\lambda}(X_1,X)\Omega^{\lambda'}(X,X_2)= \frac12\left[\delta(\lambda-\lambda')+\delta(\lambda+\lambda')\right]\Omega^{\lambda}(X_1,X_2).
\end{equation}
Here the integration measurement is defined as:
\begin{equation}
\int_{X}\equiv \int \di^{d+2}X\,\delta(X^2+1)\theta(X^0).
\end{equation}
Therefore, we can derive the inversion formulae of K\"all\'en-Lehmann representation \eqref{eq_OKL} using the orthogonality of harmonic functions, which reads:
\begin{equation}\label{InverseKL}
    \rho^{\mathcal{O}}(\wt{\nu}')=\frac1{\mathcal{N}(\wt\nu')}\int_{X_1}\Omega^{\wt{\nu}'}(X_2,X_1)D^{\mathcal{O}}_{-+}(X_1,X_2),
\end{equation}
where the normalization factor is given by:
\begin{equation}
    \mathcal{N}(\wt\nu')= \Gamma[\ii\wt\nu',-\ii\wt\nu']\Omega^{\wt\nu'}(X,X) = \frac{1}{(4\pi)^{(d+1)/2}}\Gamma\left[\begin{matrix}\frac d2+\ii\wt{\nu}',\frac d2-\ii\wt{\nu}'\\\frac{d+1}2\end{matrix}\right].
\end{equation}

\subsection{Explicit Expressions}
\paragraph{Calculation of $\rho^{\sigma^2}$}
Now we arrive at the starting point for computing the spectral functions. For the bubble function from the operator $\sigma^2$, $D^{\sigma^2}_{-+}(Y_1,Y_2) = [D^{\wt\nu}_{-+}(Y_1,Y_2)]^2$ with $\wt\nu$ being the mass parameter for $\sigma$. Therefore, the corresponding spectral function reads (after Wick rotation):
\begin{align}\label{eq_rho_sigma2}
  \rho^{\sigma^2}(\wt{\nu}')
  =&~\frac1{\mathcal{N}(\wt\nu')}\int_{X_1}\Omega^{\wt{\nu}'}(X_2,X_1)D_{-+}^{\wt{\nu}}(X_1,X_2)D_{-+}^{\wt{\nu}}(X_1,X_2)\n\\
  =&~\frac{\Gamma^2[\ii\wt\nu,-\ii\wt\nu]}{\mathcal{N}(\wt\nu')}\int_{X_1}\Omega^{\wt{\nu}'}(X_2,X_1)\Omega^{\wt{\nu}}(X_1,X_2)\Omega^{\wt{\nu}}(X_1,X_2).
\end{align}
The final task is therefore to evaluate the triple harmonic integral. The standard way to do this is to use the split representation of harmonic functions, which expresses the harmonic function as an integral of products of bulk-to-boundary propagators:
\begin{equation}\label{eq_split}
  \Omega^{\lambda}(X_1,X_2) = \frac{\lambda^2}{\pi}\int_P\Pi_{\frac d2+\ii\lambda}(X_1,P)\Pi_{\frac d2-\ii\lambda}(X_2,P) = \Omega^{-\lambda}(X_1,X_2),
\end{equation}
where $P$ denotes boundary points in the embedding space: $P^2=0$ and $P\sim \alpha P$, and $\Pi_{\Delta}(X,P)$ is the bulk-to-boundary propagator in EAdS:
\begin{equation}\label{eq_Pi}
  \Pi_{\Delta}(X,P) = \frac{1}{2\pi^{d/2}}\Gamma\left[\begin{matrix} \Delta\\ \Delta+1-\frac{d}2\end{matrix}\right](-2X\cdot P)^{-\Delta}.
\end{equation}
Also note that the integration measurement
\begin{equation}
  \int_P \equiv \frac2{\text{Vol GL}(1,\mathbb{R})^+}\int_{P^0>0}\mathrm{d}^{d+2}P\delta(P^2)
\end{equation}
reduces to $\int \di^d y$ in $\mathbb R^d$. Plugging the split representation into \eqref{eq_rho_sigma2}, we obtain:
\begin{align}
  \rho^{\sigma^2}(\wt{\nu}')=&~\frac{\wt\nu'^2\wt\nu^4\Gamma^2[\ii\wt\nu,-\ii\wt\nu]}{\pi^3\mathcal{N}(\wt\nu')}\int_{X_1}\int_{P_1,P_2,P_3}\Pi_{\frac d2-\ii\wt{\nu}'}(X_2,P_1)\Pi_{\frac d2+\ii\wt{\nu}'}(X_1,P_1)\n\\
  &\times \Pi_{\frac d2+\ii\wt{\nu}}(X_1,P_2)\Pi_{\frac d2-\ii\wt{\nu}}(X_2,P_2)\Pi_{\frac d2+\ii\wt{\nu}}(X_1,P_3)\Pi_{\frac d2-\ii\wt{\nu}}(X_2,P_3).
\end{align}
We can first perform the integration over $X_1$, which gives rise to a boundary three-point function with the standard form fixed by conformal symmetry:
\begin{equation}\label{eq_3pt}
  \int_{X_1}\Pi_{\Delta_1}(X_1,P_1)\Pi_{\Delta_2}(X_1,P_2)\Pi_{\Delta_3}(X_1,P_3) = \frac{C(\Delta_1,\Delta_2,\Delta_3)}{P_{12}^{\Delta_{123}}P_{23}^{\Delta_{231}}\,P_{31}^{\Delta_{312}}},
\end{equation}
where
\begin{equation}
  P_{ij}\equiv -2P_i\cdot P_j,\quad \Delta_{ijk}\equiv \frac{\Delta_i+\Delta_j-\Delta_k}2,
\end{equation}
and the overall coefficient is given by \cite{Freedman:1998tz,Costa:2014kfa}::
\begin{equation}
  C(\Delta_1,\Delta_2,\Delta_3) =
  \frac{1}{16\pi^d}\Gamma\left[\bgm \frac{\Delta_1+\Delta_2+\Delta_3-d}{2},\Delta_{123},\Delta_{231},\Delta_{312}\\\Delta_1+1-\frac d2,\Delta_1+2-\frac d2,\Delta_3+1-\frac d2 \edm\right].
\end{equation}
After that, we are left with integrals over boundary points $P_{1,2,3}$ of the following form: 
\begin{equation}
  I(\Delta_1,\Delta_2,\Delta_3) \equiv \int_{P_{1,2,3}}\frac{\Pi_{d-\Delta_1}(X_2,P_1)\Pi_{d-\Delta_2}(X_2,P_2)\Pi_{d-\Delta_3}(X_2,P_3)}{P_{12}^{\Delta_{123}}P_{23}^{\Delta_{231}}\,P_{31}^{\Delta_{312}}},
\end{equation}
which has been computed in \cite{Bros:2009bz,Penedones:2010ue}:
\begin{equation}
  I(\Delta_1,\Delta_2,\Delta_3) = \frac{1}{8}\Gamma\left[\bgm
  \frac d2-\Delta_{123},\frac d2-\Delta_{231},\frac d2-\Delta_{312},d-\frac{\Delta_1+\Delta_2+\Delta_3}2 \\
  d,1+\frac d2-\Delta_1,1+\frac d2-\Delta_2,1+\frac d2-\Delta_3
  \edm\right].
\end{equation}
Plugging all these together, we finally obtain:
\begin{align}\label{eq_rho_sigma2_final}
  \rho^{\sigma^2}(\wt{\nu}') =&~ \frac{\wt\nu'^2\wt\nu^4\Gamma^2[\ii\wt\nu,-\ii\wt\nu]}{\pi^3\mathcal{N}(\wt\nu')} C\left(\frac d2 +\ii\wt{\nu}',\frac d2+\ii\wt{\nu},\frac d2+\ii\wt{\nu}\right) I\left(\frac d2 +\ii\wt{\nu}',\frac d2+\ii\wt{\nu},\frac d2+\ii\wt{\nu}\right)\n\\
  =&~\FR{\wt{\nu}'\sinh(\pi\wt{\nu}')}{32\pi^{3+d/2}\Gamma[d/2,d/2+\ii\wt\nu',d/2-\ii\wt\nu']}\prod_{\aa,\bb,\cc=\pm}\Gamma\Big[ \fr d4+\fr12\ii(\aa+\bb)\wt{\nu} +\fr12\ii\cc\wt{\nu}' \Big].
\end{align}

\paragraph{Calculation of $\rho^{(\nabla\sigma)^2}$}
Similarly, for the bubble function from the operator $(\nabla\sigma)^2$, the inversion formula \eqref{eq_rho_sigma2} becomes:
\begin{align}\label{eq_rho_nabla_sigma2}
  \rho^{(\nabla\sigma)^2}(\wt{\nu}')
  =\frac{\Gamma^2[\ii\wt\nu,-\ii\wt\nu]}{\mathcal{N}(\wt\nu')}\int_{X_1}\Omega^{\wt{\nu}'}(X_2,X_1)\Big[\nabla_{X_1,A}\nabla_{X_2,B}\,\Omega^{\wt{\nu}}(X_1,X_2)\Big]^2.
\end{align}
Here the covariant derivative $\nabla$ is defined as the Wick rotation of \eqref{eq_nabla}, where the second term gains an extra minus sign:
\begin{equation}
  \nabla_A = \frac{\partial}{\partial X^A}+X_A\bigg(X\cdot\frac{\partial}{\partial X}\bigg)+ W_A\bigg(X\cdot\frac{\partial}{\partial W}\bigg),
\end{equation}
and the last term is absent for scalar fields. 
Following the same steps as the $\sigma^2$ case, we first use the split representation \eqref{eq_split}, then the derivatives acting on the harmonic functions \eqref{eq_Pi} becomes a shift in the scaling dimensions. In particular, we have:
\begin{align}
  &~\left[\nabla_{X} \Pi_{\Delta}(X,P_2)\right]\cdot \left[\nabla_{X} \Pi_{\Delta}(X,P_3)\right] = \Pi_{\Delta}(X,P_2)\Pi_{\Delta}(X,P_3) \left( \Delta X + \frac{\Delta}{X\cdot P_2} P_2 \right) \cdot \left( \Delta X + \frac{\Delta}{X\cdot P_3} P_3 \right)\n\\
  &\qquad\qquad = \Delta^2\Pi_\Delta(X,P_2)\Pi_\Delta(X,P_3)+(2\Delta+2-d)^2(P_2\cdot P_3)\Pi_{\Delta+1}(X,P_2)\Pi_{\Delta+1}(X,P_3).
\end{align}
Therefore, the spectral function \eqref{eq_rho_nabla_sigma2} becomes:
\begin{align}
  \rho^{(\nabla \sigma)^2}(\wt{\nu}')=&~\frac{\wt\nu'^2\wt\nu^4\Gamma^2[\ii\wt\nu,-\ii\wt\nu]}{\pi^3\mathcal{N}(\wt\nu')}\int_{X_1}\int_{P_1,P_2,P_3}\Pi_{\frac d2-\ii\wt{\nu}'}(X_2,P_1)\Pi_{\frac d2+\ii\wt{\nu}'}(X_1,P_1)\n\\
  &\times \left[\nabla_{X_1}\Pi_{\frac d2+\ii\wt{\nu}}(X_1,P_2)\cdot\nabla_{X_1}\Pi_{\frac d2+\ii\wt{\nu}}(X_1,P_3)\right]\left[\nabla_{X_2}\Pi_{\frac d2-\ii\wt{\nu}}(X_2,P_2)\cdot\nabla_{X_2}\Pi_{\frac d2-\ii\wt{\nu}}(X_2,P_3)\right] \n\\
  =&~\frac{\wt\nu'^2\wt\nu^4\Gamma^2[\ii\wt\nu,-\ii\wt\nu]}{\pi^3\mathcal{N}(\wt\nu')}\int_{X_1}\int_{P_1,P_2,P_3}\Pi_{\frac d2-\ii\wt{\nu}'}(X_2,P_1)\Pi_{\frac d2+\ii\wt{\nu}'}(X_1,P_1)\n\\
  &\times \left[(\frac d2+\ii\wt\nu)^2\Pi_{\frac d2+\ii\wt{\nu}}(X_1,P_2)\Pi_{\frac d2+\ii\wt{\nu}}(X_1,P_3)-2(1+\ii\wt\nu)^2 P_{23}\Pi_{\frac d2+1+\ii\wt{\nu}}(X_1,P_2)\Pi_{\frac d2+1+\ii\wt{\nu}}(X_1,P_3)\right]\n\\
  &\times \left[(\frac d2-\ii\wt\nu)^2\Pi_{\frac d2-\ii\wt{\nu}}(X_2,P_2)\Pi_{\frac d2-\ii\wt{\nu}}(X_2,P_3)-2(1-\ii\wt\nu)^2 P_{23}\Pi_{\frac d2+1-\ii\wt{\nu}}(X_2,P_2)\Pi_{\frac d2+1-\ii\wt{\nu}}(X_2,P_3)\right]\n\\
  =&~\frac{\wt\nu'^2\wt\nu^4\Gamma^2[\ii\wt\nu,-\ii\wt\nu]}{\pi^3\mathcal{N}(\wt\nu')}\bigg[
  (\frac{d^2}4+\wt\nu^2)^2\, C\left(\frac d2+\ii\wt\nu',\frac d2+\ii\wt\nu,\frac d2+\ii\wt\nu\right) I\left(\frac d2+\ii\wt\nu',\frac d2+\ii\wt\nu,\frac d2+\ii\wt\nu\right)\n\\
  &-2(\frac d2+\ii\wt\nu)^2(1-\ii\wt\nu)^2\, C\left(\frac d2+\ii\wt\nu',\frac d2+\ii\wt\nu,\frac d2+\ii\wt\nu\right) I\left(\frac d2+\ii\wt\nu',\frac d2-1+\ii\wt\nu,\frac d2-1+\ii\wt\nu\right) \n\\  
  &-2(\frac d2-\ii\wt\nu)^2(1+\ii\wt\nu)^2\, C\left(\frac d2+1+\ii\wt\nu',\frac d2+1+\ii\wt\nu,\frac d2+\ii\wt\nu\right) I\left(\frac d2+\ii\wt\nu',\frac d2+\ii\wt\nu,\frac d2+\ii\wt\nu\right) \n\\
  &+4(1+\wt\nu^2)^2\, C\left(\frac d2+\ii\wt\nu',\frac d2+1+\ii\wt\nu,\frac d2+1+\ii\wt\nu\right) I\left(\frac d2+\ii\wt\nu',\frac d2-1+\ii\wt\nu,\frac d2-1+\ii\wt\nu\right)\bigg] \n\\
  =&~\left(\wt\nu^2-\frac{\wt\nu'^2}{2}+\frac{d^2}{8}\right)^2\rho^{\sigma^2}(\wt\nu'),
\end{align}
where $\rho^{\sigma^2}(\wt\nu')$ is given in \eqref{eq_rho_sigma2_final}.

\paragraph{Calculation of $\rho^{A^2}$}
To deal with vector fields, we should introduce the split representation for the harmonic functions with spin \cite{Costa:2014kfa}: 
\begin{equation}\label{eq_splitA}
  \Omega^{\lambda}_\ell(X_1,X_2;W_1,W_2) = \frac{\lambda^2}{\pi\ell!\big(\frac d2-1\big)_\ell}\int_P\Pi_{\frac d2+\ii\lambda,\ell}(X_1,P;W_1,D_Z)\Pi_{\frac d2-\ii\lambda,\ell}(X_2,P;W_2,Z).
\end{equation}
Here $Z$ is an auxiliary vector and $D_Z$ is defined to be
\begin{equation}
  D_Z^A=\bigg(\frac d2-1+Z\cdot\frac{\partial}{\partial Z}\bigg)\frac{\partial}{\partial Z_A}-\frac12Z^A\frac{\partial^2}{\partial Z\cdot\partial Z},
\end{equation}
and the spinning bulk-to-boundary propagator is given by
\begin{equation}
  \Pi_{\Delta,\ell}(X,P;W,Z) = \frac{(\ell+\Delta-1)\Gamma(\Delta)}{2\pi^{\frac d2}(\Delta-1)\Gamma\big(\Delta+1-\frac d2\big)}\frac{((-2P\cdot X)(W\cdot Z)+2(W\cdot P)(Z\cdot X))^\ell}{(-2P\cdot X)^{\Delta+\ell}}.
\end{equation}
Therefore, the spectral function $\rho^{A^2}(\wt\nu')$ can be computed via the inversion formula \eqref{InverseKL}:
\begin{align}\label{eq_rhoA2}
  \rho^{A^2}(\wt\nu')=&~\frac{\Gamma^2[\ii\wt\nu_A,-\ii\wt\nu_A]}{\mathcal{N}(\wt\nu')}\int_{X_1}\Omega^{\wt{\nu}'}(X_2,X_1)\left[K_{1,A}K_{2,B}\,\Omega^{\wt\nu_A}_1(X_1,X_2;W_1,W_2)\right]^2,
\end{align}
where $\wt\nu_A$ denotes the mass parameter for the vector field $A$, and $K$ is the operator in \eqref{KVector} after Wick rotation:
\begin{align}
    K_A = &~\bigg(\frac{d-1}2\bigg)\bigg[\frac{\partial}{\partial W^A}+X_A\bigg(X\cdot\frac{\partial}{\partial W}\bigg)\bigg]+\bigg(W\cdot\frac{\partial}{\partial W}\bigg)\frac{\partial}{\partial W^A}+X_A\bigg(X\cdot\frac{\partial}{\partial W}\bigg)\bigg(W\cdot\frac{\partial}{\partial W}\bigg)\n \\
    &-\frac12W_A\bigg[\frac{\partial^2}{\partial W\cdot\partial W}+\bigg(X\cdot\frac{\partial}{\partial W}\bigg)\bigg(X\cdot\frac{\partial}{\partial W}\bigg)\bigg].
\end{align}
Substituting the split representation \eqref{eq_splitA} into \eqref{eq_rhoA2}, we obtain:
\begin{align}
  \rho^{A^2}(\wt\nu')=&~\frac{\wt\nu'^2\wt\nu_A^4\Gamma^2[\ii\wt\nu_A,-\ii\wt\nu_A]}{\pi^3(d/2-1)^2\mathcal{N}(\wt\nu')}\int_{X_1}\int_{P_1,P_2,P_3}\Pi_{\frac d2-\ii\wt{\nu}'}(X_2,P_1)\Pi_{\frac d2+\ii\wt{\nu}'}(X_1,P_1)\n\\
  &\times \left[K_{1}\Pi_{\frac d2+\ii\wt{\nu}_A,1}(X_1,P_2;W_1,D_{Z_1})\cdot K_{1}\Pi_{\frac d2+\ii\wt{\nu}_A,1}(X_1,P_3;W_1,D_{Z_2})\right]\n\\
  &\times \left[K_{2}\Pi_{\frac d2-\ii\wt{\nu}_A,1}(X_2,P_2;W_2,Z_1)\cdot K_{2}\Pi_{\frac d2-\ii\wt{\nu}_A,1}(X_2,P_3;W_2,Z_2)\right].
\end{align}
It is not hard to check:
\begin{align}\label{eq_KPi}
  K \Pi_{\Delta,1}(X,P;W,Z) =&~ \frac{(d-1)\Gamma(\Delta+1)}{2\pi^{\frac d2}(\Delta-1)\Gamma\big(\Delta+1-\frac d2\big)}\frac{(Z\cdot X)P-(P\cdot X)Z}{(-2P\cdot X)^{\Delta+1}}\n\\
  =&~\frac{d-1}{2(\Delta-1)}\left[P(Z\cdot\partial_P)+\Delta Z\right]\Pi_\Delta(X,P).
\end{align}
Therefore, the integration over $X_1$ in \eqref{eq_rhoA2} is in the same form of \eqref{eq_3pt} and thus doable. Let $\Delta_A \equiv d/2+\ii\wt\nu_A$ and $\Delta'\equiv d/2+\ii\wt\nu'$ for simplicity, we have:
\begin{align}\label{eq_temp2}
  &~\left[P_2(D_{Z_1}\cdot\partial_{P_2})+\Delta_A D_{Z_1}\right]\cdot
  \left[P_3(D_{Z_2}\cdot\partial_{P_3})+\Delta_A D_{Z_2}\right]\int_{X_1}\Pi_{\Delta'}(X_1,P_1)\Pi_{\Delta_A}(X_1,P_2)\Pi_{\Delta_A}(X_1,P_3)\n\\
  =&~\left[P_2(D_{Z_1}\cdot\partial_{P_2})+\Delta_A D_{Z_1}\right]\cdot
  \left[P_3(D_{Z_2}\cdot\partial_{P_3})+\Delta_A D_{Z_2}\right]
  \frac{C(\Delta',\Delta_A,\Delta_A)}{P_{12}^{\Delta'/2}P_{23}^{\Delta_A-\Delta'/2}\,P_{31}^{\Delta'/2}}\n\\
  =&~\frac{C(\Delta',\Delta_A,\Delta_A)}{P_{12}^{\Delta'/2}P_{23}^{\Delta_A-\Delta'/2}P_{31}^{\Delta'/2}}\bigg[(\Delta_A^2-\Delta_A+\frac{\Delta'}{2})(D_{Z_1}\cdot D_{Z_2})+\frac{(2\Delta_A-\Delta')(\Delta_A+\frac{\Delta'}{2}-1)}{P_{23}}(D_{Z_1}\cdot P_3)(D_{Z_2}\cdot P_2)\n\\
  &+\frac{\Delta'^2}{2P_{12}}(D_{Z_1}\cdot P_1)(D_{Z_2}\cdot P_2)+\frac{\Delta'^2}{2P_{31}}(D_{Z_1}\cdot P_3)(D_{Z_2}\cdot P_1)-\frac{\Delta'^2P_{23}}{2P_{12}P_{31}}(D_{Z_1}\cdot P_1)(D_{Z_2}\cdot P_1)\bigg].
\end{align}
Plugging \eqref{eq_temp2} and \eqref{eq_KPi} into \eqref{eq_rhoA2}, we finally obtain:
\begin{align}\label{eq_temp3}
  \rho^{A^2}(\wt\nu')=&~\frac{\wt\nu'^2\wt\nu_A^4\Gamma^2[\ii\wt\nu_A,-\ii\wt\nu_A]}{\pi^3(d/2-1)^2\mathcal{N}(\wt\nu')}\frac{(d-1)^4(d-\Delta_A)^2 }{16(\Delta_A-1)^2(d-1-\Delta_A)^2}\int_{P_1,P_2,P_3}\frac{C(\Delta',\Delta_A,\Delta_A)\Pi_{d-\Delta'}(X_2,P_1)}{P_{12}^{\Delta'/2}P_{23}^{\Delta_A-\Delta'/2}P_{31}^{\Delta'/2}}\n\\
  &\times \bigg[(\Delta_A^2-\Delta_A+\frac{\Delta'}{2})(D_{Z_1}\cdot D_{Z_2})+\frac{(2\Delta_A-\Delta')(\Delta_A+\frac{\Delta'}{2}-1)}{P_{23}}(D_{Z_1}\cdot P_3)(D_{Z_2}\cdot P_2)\n\\
  &+\frac{\Delta'^2}{2P_{12}}(D_{Z_1}\cdot P_1)(D_{Z_2}\cdot P_2)+\frac{\Delta'^2}{2P_{31}}(D_{Z_1}\cdot P_3)(D_{Z_2}\cdot P_1)-\frac{\Delta'^2P_{23}}{2P_{12}P_{31}}(D_{Z_1}\cdot P_1)(D_{Z_2}\cdot P_1)\bigg]\n\\
  &\times \left(Z_1-\frac{Z_1\cdot X_2}{P_2\cdot X_2}P_2\right)\cdot \left(Z_2-\frac{Z_2\cdot X_2}{P_3\cdot X_2}P_3\right)\Pi_{d-\Delta_A}(X_2,P_2)\Pi_{d-\Delta_A}(X_2,P_3).
\end{align}

Since it is linear in $Z_1$ and $Z_2$, the action of $D_{Z_{1,2},A}$ on \eqref{eq_temp3} simply replaces $Z_{1,2}\cdot Q$ with $(d/2-1)Q_A$. After simplification, we can express the spectral function $\rho^{A^2}(\wt\nu')$ as a linear combination of integrals of the following form:
\begin{align}
  \rho^{A^2}(\wt\nu')=&~\frac{\wt\nu'^2\wt\nu_A^4\Gamma^2[\ii\wt\nu_A,-\ii\wt\nu_A]}{\pi^3\mathcal{N}(\wt\nu')}\frac{(d-1)^4(d-\Delta_A)^2 }{16(\Delta_A-1)^2(d-1-\Delta_A)^2}\int_{P_1,P_2,P_3}\frac{C(\Delta',\Delta_A,\Delta_A)}{P_{12}^{\Delta'/2}P_{23}^{\Delta_A-\Delta'/2}P_{31}^{\Delta'/2}}\n\\
  &\times \bigg[(d-2)(\Delta_A^2-\Delta_A+\frac{\Delta'}{2})+\frac{\Delta'^2P_{31}}{4P_{12}}\frac{P_2\cdot X_2}{P_3\cdot X_2}+\frac{\Delta'^2P_{12}}{4P_{31}}\frac{P_3\cdot X_2}{P_2\cdot X_2}\n\\
  &-\frac{\Delta'^2P_{23}}{2P_{12}}\frac{P_1\cdot X_2}{P_3\cdot X_2}-\frac{\Delta'^2P_{23}}{2P_{31}}\frac{P_1\cdot X_2}{P_2\cdot X_2}+\frac{\Delta'^2P_{23}^2}{4P_{12}P_{31}}\frac{(P_1\cdot X_2)^2}{(P_2\cdot X_2)(P_3\cdot X_2)}\n\\
  &+\frac{(2\Delta_A^2-2\Delta_A+\Delta')P_{23}}{4(P_2\cdot X_2)(P_3\cdot X_2)}\bigg]\Pi_{d-\Delta'}(X_2,P_1)\Pi_{d-\Delta_A}(X_2,P_2)\Pi_{d-\Delta_A}(X_2,P_3) \n\\
  =&~\frac{\wt\nu'^2\wt\nu_A^4\Gamma^2[\ii\wt\nu,-\ii\wt\nu]}{\pi^3\mathcal{N}(\wt\nu')}\frac{(d-1)^4(d-\Delta_A)^2 }{16(\Delta_A-1)^2(d-1-\Delta_A)^2}C(\Delta',\Delta_A,\Delta_A)\n\\
  &\times \bigg\{(d-2)(\Delta_A^2-\Delta_A+\frac{\Delta'}{2}) I(\Delta',\Delta_A,\Delta_A)\n\\
  &+\frac{\Delta'^2(d+2-2\Delta_A)(d-1-\Delta_A)}{4(d-2\Delta_A)(d-\Delta_A)}
  \Big[ I(\Delta',\Delta_A-1,\Delta_A+1)+ I(\Delta',\Delta_A+1,\Delta_A-1)\Big]\n\\
  &-\frac{\Delta'^2(d+2-2\Delta_A)(d-1-\Delta')}{2(d-2\Delta')(d-\Delta_A)}
  \Big[ I(\Delta'+1,\Delta_A-1,\Delta_A)+ I(\Delta'+1,\Delta_A,\Delta_A-1)\Big]\n\\
  &+\frac{\Delta'^2(d+2-2\Delta_A)^2(d-1-\Delta')(d-2-\Delta')}{4(d-2\Delta')(d-2-2\Delta')(d-\Delta_A)^2} I(\Delta'+2,\Delta_A-1,\Delta_A-1)\n\\
  &+\frac{(2\Delta_A^2-2\Delta_A+\Delta')(d+2-2\Delta_A)^2}{4(d-\Delta_A)^2} I(\Delta',\Delta_A-1,\Delta_A-1)\bigg\}\n\\
  =&~\frac{(d-1)^4}{64[4\wt\nu_A^2+(d-2)^2]^2}\times \Big[64d\wt\nu_A^4-64\wt\nu_A^2\wt\nu'^2+16\wt\nu'^4 + 16d(2d^2-5d+4)\wt\nu_A^2 \n\\
  &+ 8d(3d-4)\wt\nu'^2+(4d^2-11d+8)d^3\Big]\rho^{\sigma_{\wt\nu_A}^2}(\wt\nu'),
\end{align}
where $\rho^{\sigma_{\wt\nu_A}^2}(\wt\nu')$ is the spectral density of the $\sigma^2$ bubble in \eqref{eq_rho_sigma2_final} with the mass parameter of $\sigma$ replaced by $\wt\nu_A$.

\end{appendix}

\newpage
\bibliography{CosmoCollider} 
\bibliographystyle{utphys}

\end{document}